\documentclass[twocolumn,aps,amsmath,prd,amssymb,eqsecnum,nofootinbib]{revtex4}

\usepackage{graphicx}

\begin{document}

\title{Static Kerr Green's Function in Closed Form and an Analytic Derivation of the Self-Force for a Static Scalar Charge in Kerr Space-Time }
\author{Adrian C. Ottewill}
\email{adrian.ottewill@ucd.ie}
\affiliation{School of Mathematical Sciences and Complex \& Adaptive Systems Laboratory, University College Dublin, Belfield, Dublin 4, Ireland}
\author{Peter Taylor}
\email{peter.taylor@maths.tcd.ie}
\affiliation{School of Mathematics, Trinity College, Dublin 2, Ireland}

\date{\today}

\begin{abstract}
We derive a closed-form solution for the Green's function for the wave equation of a static (with respect to an undragged, static observer at infinity) scalar charge in the Kerr space-time. We employ our solution to obtain an analytic expression for the self-force on such a charge, comparing our results to those of Ref.~\cite{BurkoLiu}.~\end{abstract}
\maketitle

\section{Introduction}
\label{sec:intro}
Understanding the Green's function on the Kerr geometry is important in a number of fundamental theoretical problems such as the radiation reaction problem \cite{Barack2010, Barack2011}, miniature black hole creation \cite{Winstanley2008} and in the Kerr/CFT correspondence \cite{Strominger1, Strominger2, Strominger3}. The wave equations satisfied by Green's functions for black hole space-times are extremely complicated, even in the most simple case of a Schwarzschild black hole. However, in the Schwarzschild case,  despite not having closed-form solutions for the Green's functions, the static or zero-frequency scalar Green's function is known in closed form and may be used to obtain, for example, analytic expressions for the vacuum polarization on the horizon of the black hole \cite{Candelas:1980zt} or the self-force on a static scalar charge \cite{Wiseman}.

For the Kerr space-time, as in the Schwarzschild case, the wave equation is fully separable \cite{Teukolsky} and we may therefore obtain a mode-sum representation for the Green's function in terms of spheroidal functions and complicated radial functions that must be computed numerically. The zero-frequency mode simplifies significantly though still a highly non-trivial double mode-sum over a product of four associated Legendre functions, not all of integer order. The only closed-form expression for a Green's function was obtained by Linet \cite{LinetPoleKerr}, where he derived the scalar Green's function with one point on the pole of the Kerr black hole and Frolov \cite{Frolov:1982pi} used this result to calculate analytically the vacuum polarization on the pole of the Kerr black hole.

In this paper, we shall obtain the closed-form representation of the Green's function for a static scalar charge in the Kerr space-time, i.e., a scalar charge at fixed spatial Boyer-Lindquist coordinates. This derivation relies on two interesting results concerning the associated Legendre functions, one of which we prove in this paper and the other we prove in Ref.~\cite{CasalsOttewillTaylor}. We validate our expression for the Green's function by calculating the self-force on a static scalar charge in the Kerr space-time, obtaining the result conjectured by Burko and Liu \cite{BurkoLiu}. Therefore, this paper is also the first proof of the result claimed in Ref.~\cite{BurkoLiu}. 


\section{The Kerr Green's Function}
\label{sec:modesumgreensfn}
The Kerr metric in Boyer-Lindquist coordinates is
\begin{align}
ds^{2}=-\frac{(\Delta-a^{2}\sin^{2}\theta)}{\Sigma}dt^{2}-2a\sin^{2}\theta\frac{(r^{2}+a^{2}-\Delta)}{\Sigma}dt d\phi\nonumber\\ +\Big(\frac{(r^{2}+a^{2})^{2}-a^{2}\Delta\sin^{2}\theta}{\Sigma}\Big)\sin^{2}\theta d\phi^{2}
+\frac{\Sigma}{\Delta} dr^{2}+\Sigma d\theta^{2}
\end{align}
where
\begin{equation}
\Delta = r^{2}+a^{2}-2Mr\qquad\textrm{and}\qquad \Sigma=r^{2}+a^{2}\cos^{2}\theta.
\end{equation}
The event and Cauchy horizons are defined by vanishing $\Delta$, which has two roots $r_{\pm}=M\pm\sqrt{M^{2}-a^{2}}$, the positive subscript being the event horizon and the negative being the Cauchy horizon. Moreover, the Killing vector $\partial/\partial t$ is the null generator of the static limit surface, given by $r_{e}=M+\sqrt{M^{2}-a^{2}\cos^{2}\theta}$. Inside this surface, the ergosphere region, $\partial/\partial t$ becomes space-like and so local radial null cones tip over forcing particle's to move in the direction of the black hole's rotation, i.e., a particle inside the static limit surface cannot remain static. We would therefore expect our closed-form expression for the static scalar Green's function to diverge as the static limit surface is approached.

The Green's function for a scalar particle in the Kerr space-time satisfies the following wave equation:
\begin{align}
\label{eq:waveeqnkerr}
\Big\{-\Big(\frac{(r^{2}+a^{2})^{2}}{\Delta}-a^{2}\sin^{2}\theta\Big)\frac{\partial^{2}}{\partial t^{2}}-\frac{4 Ma r}{\Delta}\frac{\partial^{2}}{\partial t\partial\phi}\nonumber\\
+\Big(\frac{1}{\sin^{2}\theta}-\frac{a^{2}}{\Delta}\Big)\frac{\partial^{2}}{\partial\phi^{2}}+\frac{1}{\sin\theta}\frac{\partial}{\partial \theta}\Big(\sin\theta\frac{\partial}{\partial \theta}\Big)\nonumber\\
+\frac{\partial}{\partial r}\Big(\Delta\frac{\partial}{\partial r}\Big)\Big\}G(x,x')=-\frac{\delta(x-x')}{\sin\theta}.
\end{align}
The wave equation on this metric admits a fully separable solution \cite{Teukolsky} and so the Green's function may be written as the following mode-sum representation in terms of separated mode-solutions:
\begin{align}
\label{eq:modesumfull}
G(x,x')=&\frac{1}{2\pi^{2}}\int_{0}^{\infty}\cos\omega(t-t')\sum_{l=0}^{\infty}\sum_{m=-l}^{l}e^{im(\phi-\phi')}\nonumber\\
&S_{\omega lm}(\cos\theta)S_{\omega lm}(\cos\theta') \chi_{\omega lm}(r,r')d\omega
\end{align}
where $S_{\omega lm}(\cos\theta)$ are the normalized spheroidal functions satisfying
\begin{align}
\label{eq:spheroidal}
\Big\{\frac{1}{\sin\theta}\frac{d}{d\theta}\Big(\sin\theta\frac{d}{d\theta}\Big)+a^{2}\omega^{2}\cos^{2}\theta-\frac{m^{2}}{\sin^{2}\theta}\nonumber\\
+\lambda_{\omega lm}\Big\}S_{\omega l m}(\cos\theta)=0
\end{align}
and also satisfying the normalization condition
\begin{equation}
\label{eq:normalization}
\int_{-1}^{1}S_{\omega lm}(\cos\theta) S_{\omega l' m}(\cos\theta) \,\,d(\cos\theta)=\delta_{l l'}.
\end{equation}
Eq.~(\ref{eq:spheroidal}) and the normalization condition (\ref{eq:normalization}) imply that for $\omega=0$, the spheroidal functions reduce to the normalized associated Legendre polynomials, i.e.,
\begin{align}
S_{0lm}(\cos\theta)=\sqrt{\frac{(2l+1)}{2}\frac{(l-m)!}{(l+m)!}}P^{m}_{l}(\cos\theta).
\end{align}
The radial part of the Green's function, $\chi_{\omega lm}$, satisfies the inhomogeneous equation
\begin{align}
\Big\{\frac{d}{dr}\Big(\Delta\frac{d}{dr}\Big)+\omega^{2}\frac{(r^{2}+a^{2})^{2}}{\Delta}+\frac{4 M ar m\omega}{\Delta}+\frac{m^{2} a^{2}}{\Delta}\nonumber\\
-a^{2}\omega^{2}-\lambda_{\omega lm}\Big\}\chi_{\omega l m}(r,r')=-\delta(r-r').
\end{align}
For a static source, the derivatives with respect to $t$ vanish in Eq.~(\ref{eq:waveeqnkerr}). Hence the static Green's function satisfies
\begin{align}
\label{eq:waveeqstatic}
\Big\{\Big(\frac{1}{\sin^{2}\theta}-\frac{a^{2}}{\Delta}\Big)\frac{\partial^{2}}{\partial\phi^{2}}+\frac{1}{\sin\theta}\frac{\partial}{\partial \theta}\Big(\sin\theta\frac{\partial}{\partial \theta}\Big)\nonumber\\
+\frac{\partial}{\partial r}\Big(\Delta\frac{\partial}{\partial r}\Big)\Big\}G_{\textrm{static}}(\textbf{x},\textbf{x}')=-\frac{\delta(\textbf{x}-\textbf{x}')}{\sin\theta}.
\end{align}
In this case, the mode-sum expression reduces to the zero-frequency mode of Eq.~(\ref{eq:modesumfull}) (modulo a factor of $2\pi$) resulting in a simplified mode-sum solution in terms of associated Legendre functions
\begin{align}
\label{eq:statickerrmodesum}
G_{\textrm{static}}(\textbf{x},\textbf{x}')=&\frac{1}{4\pi}\sum_{l=0}^{\infty}\sum_{m=-l}^{l}e^{im(\phi-\phi')}(2l+1)\frac{(l-m)!}{(l+m)!} \nonumber\\
&P_{ l}^{m}(\cos\theta)P_{l}^{m}(\cos\theta') \chi_{ lm}(r,r')
\end{align}
where $\chi_{lm}(r,r')$ satisfies
\begin{align}
\Big\{\frac{d}{dr}\Big(\Delta\frac{d}{dr}\Big)+\frac{m^{2} a^{2}}{\Delta}-l(l+1)\Big\}\chi_{ l m}(r,r')=-\delta(r-r').
\end{align}
Changing the radial variable
\begin{equation}
\eta=\frac{r-M}{b}\qquad\textrm{where}\quad b=\sqrt{M^{2}-a^{2}},
\end{equation}
we may rewrite the radial equation as
\begin{align}
\label{eq:radialeqneta}
\Big\{\frac{d}{d\eta}\Big((\eta^{2}-1)\frac{d}{d\eta}\Big)+\frac{\gamma^{2}m^{2}}{(\eta^{2}-1)}&-l(l+1)\Big\}\chi_{ lm}(\eta,\eta')\nonumber\\
&=-\frac{\delta(\eta-\eta')}{b},
\end{align}
where $\gamma=a/b$. We note that the regular singular point has now been shifted from $r=r_{+}$ to $\eta=1$. The solutions of the corresponding homogeneous equation are the associated Legendre functions of pure imaginary, non-integer order, $P_{l}^{\pm i \gamma m}(\eta)$ and $Q_{l}^{\pm i \gamma m}(\eta)$. The boundary conditions imposed correspond to different Green's functions which in turn correspond to different combinations of these associated Legendre functions. For the retarded Green's function, for example, one typically chooses ingoing radiation boundary conditions at the horizon and outgoing radiation boundary conditions at infinity. However, since we are considering a static particle, we cannot impose radiation boundary conditions. Rather, for the retarded Green's function, we require that the inner solution be regular on the future event horizon in ingoing Kerr-Newman coordinates while the advanced Green's function corresponds to the inner solution being regular on the past event horizon in outgoing Kerr-Newman coordinates. Near the horizon, the two candidates for the inner solution, $P_{l}^{\pm i \gamma m}(\eta)$, are oscillatory
\begin{equation}
P_{l}^{\pm i\gamma m}(\eta)\sim \frac{1}{\Gamma(1\mp i \gamma m)}\left(\frac{\eta+1}{\eta-1}\right)^{\pm i \gamma m/2}\quad \textrm{as}\qquad \eta\rightarrow1^{+}.
\end{equation}

In ingoing Kerr-Newman coordinates $(r, \theta, \tilde{\phi})$, where $\tilde{\phi}$ is related to Boyer-Lindquist $\phi$ by
\begin{equation}
d\tilde{\phi}=d\phi+\frac{a}{\Delta}d r,
\end{equation}
it may be shown that
\begin{align}
P_{l}^{i m \gamma}(\eta) Y_{lm}(\theta, \phi)&\sim  \frac{1}{\Gamma(1-i m \gamma)}Y_{lm}(\theta, \tilde{\phi})\left(\frac{\eta+1}{\eta-1}\right)^{i m \gamma}\nonumber\\
 P_{l}^{-i m \gamma}(\eta) Y_{lm}(\theta, \phi)& \sim  \frac{1}{\Gamma(1+i m \gamma)}Y_{lm}(\theta, \tilde{\phi}),
\end{align}
as $\eta\rightarrow 1^{+}$. It is clear from these asymptotic expressions, that the first solution oscillates infinitely fast in ingoing Kerr-Newman coordinates whereas the second solution is regular. The appropriate inner solution for the retarded Green's function is therefore $P_{l}^{-i m \gamma}(\eta)$. Similarly, for the advanced Green's function, the appropriate choice is $P_{l}^{im\gamma}(\eta)$.

Having fixed our choice for the inner solution, the choice for the outer solution is determined by regularity at infinity; we may choose either $Q_{l}^{i m\gamma}(\eta)$ or $Q_{l}^{-i m \gamma}$ since both are regular at infinity and differ only by a constant multiplier for each $l$ and $m$. This constant multiplier will be accounted for by the normalization. The solution of the retarded inhomogeneous radial equation Eq.(\ref{eq:radialeqneta}) may be written formally as
\begin{equation}
\chi_{lm}(\eta, \eta')=\frac{1}{b}\frac{P_{l}^{-\nu} (\eta_{<})Q_{l}^{\nu}(\eta_{>})}{N}=\frac{1}{b}\frac{P_{l}^{-\nu} (\eta_{<})Q_{l}^{-\nu}(\eta_{>})}{\bar{N}},
\end{equation}
where $N=-(z^{2}-1)W[P^{-\nu}_{l}(z),Q^{\nu}_{l}(z)]$ and $\bar{N}=-(z^{2}-1)W[P^{-\nu}_{l}(z), Q^{-\nu}_{l}(z)]$ are constant and $W[P, Q]$ is the Wronskian. For the pair of solutions $\big\{P_{l}^{-\nu}(\eta), Q_{l}^{\nu}(\eta)\big\}$, the Wrosnkian takes a particularly simple form \cite{gradriz}:
\begin{align}
W[P^{-\nu}_{l}, Q^{\nu}_{l}]&=P^{-\nu}_{l}(z)\frac{d Q^{\nu}_{l}(z)}{d z}-\frac{d P^{-\nu}_{l}(z)}{d z} Q^{\nu}_{l}(z)\nonumber\\
&=-\frac{e^{i \nu\pi}}{(z^{2}-1)}
\end{align}
and so
\begin{equation}
\chi_{lm}(\eta, \eta')=\frac{1}{b} e^{-i \nu \pi} P_{l}^{-\nu}(\eta_{<}) Q_{l}^{\nu}(\eta_{>}) \quad \textrm{where}\quad \nu=i m\gamma.
\end{equation}
The final form of the static retarded Green's function is therefore
\begin{align}
\label{eq:greensfnmodesum}
G_{\textrm{static}}^{\textrm{ret}}(\textbf{x},\textbf{x}')=\frac{1}{4\pi b}\sum_{l=0}^{\infty}\sum_{m=-l}^{l}e^{-im\Delta\phi}(2l+1)\frac{(l-m)!}{(l+m)!}\nonumber\\
 P_{ l}^{m}(\cos\theta)P_{l}^{m}(\cos\theta')\, e^{-i \nu \pi} P_{l}^{-\nu}(\eta_{<}) Q_{l}^{\nu}(\eta_{>})
\end{align}
where $\Delta\phi=\phi'-\phi$.

The form of this Green's function looks similar to those that are known in closed-form. In particular, considering the three-dimensional Euclidean Green's function in spheroidal coordinates gives us a product of four Legendre functions of integer order similar to the mode-sum above. In \cite{OttewillTaylor1}, we have shown how one may also obtain a quasi-closed form for a product of four Legendre functions of non-integer order and degree by considering the Green's function on a three-dimensional cosmic string space-time in spheroidal coordinates. In the static Kerr case however, one has both integer and non-integer (pure imaginary) order Legendre functions and it is precisely this mixing that has made this problem so elusive for so long. In the following sections we shall show that the mode-sum above may indeed be performed leading to a completely closed-form expression for the static Kerr Green's function. In this paper, we offer one particularly neat application of this closed-form Green's function, where we obtain an analytic expression for the self-force on a static scalar charge in the Kerr space-time and we compare our result to the conjectured result of  Ref.~\cite{BurkoLiu}. We note also that the result of Ref.~\cite{BurkoLiu} relies on a conjecture for the MSRP parameter $D_{\mu}$ which has recently been computed, along with higher order parameters, for the Schwarzschild \cite{Heffernan} and Kerr \cite{Heffernan2} space-times.

\section{The Static Kerr Green's Function In Closed Form}
\label{sec:closedformgreensfn}
In this section, we derive the static Kerr Green's function in closed form for the field satisfying retarded boundary conditions; the derivation for the advanced field is analogous. Obtaining the closed-form solution relies crucially on two formulae involving the associated Legendre functions, results which may well prove useful in other contexts. 

In Ref.~\cite{CasalsOttewillTaylor}, it is shown that a product of Associated Legendre functions of arbitrary order may be written in the following integral form:
\begin{align}
\label{eq:brproduct}
&e^{-i\nu\pi}P_{l}^{-\nu}(\eta_{<})Q_{l}^{\nu}(\eta_{>})\nonumber\\
&=\frac{1}{2}\int_{-1}^{1}\frac{e^{-\nu\cosh^{-1}(\chi)}P_{l}(x)}{(\eta^{2}+\eta'^{2}-2\eta\eta' x-1+x^{2})^{1/2}}dx
\end{align}
where
\begin{equation}
\chi=\frac{\eta\eta'-x}{(\eta^{2}-1)^{1/2}(\eta'^{2}-1)^{1/2}}.
\end{equation}
This is a generalization of  a result obtained in Ref.~\cite{Candelas:1984pg} for a product of Associated Legendre functions of integer order. The derivation in \cite{CasalsOttewillTaylor} involves obtaining a closed-form solution for the wave-equation on a dimensionally reduced Bertotti-Robinson space-time and equating with the equivalent mode-sum expression. Employing our result (\ref{eq:brproduct}) in our expression for the Green's function (\ref{eq:greensfnmodesum}), we obtain
\begin{align}
G_{\textrm{static}}(\textbf{x},\textbf{x}')=\frac{1}{8\pi b}\int_{-1}^{1}\sum_{m=-\infty}^{\infty}\sum_{l=|m|}^{\infty}e^{-im(\Delta\phi+\gamma\cosh^{-1}(\chi))}\nonumber\\
(2l+1)\frac{(l-m)!}{(l+m)!}\frac{ P_{ l}^{m}(\cos\theta')P_{l}^{m}(\cos\theta)P_{l}(x)}{(\eta^{2}+\eta'^{2}-2\eta\eta' x -1+x^{2})^{1/2}}\, dx .
\end{align}
where we have changed the order of summation in the expression above. 

In the Appendix, we derive the following summation formula for the product of three Legendre functions
\begin{align}
\label{eq:baranovsum}
\sum_{l=|m|}^{\infty}(2l+1)\frac{(l-m)!}{(l+m)!}P_{l}^{m}(\cos\theta)P_{l}^{m}(\cos\theta')P_{l}(x)\nonumber\\
=\frac{2}{\pi}\frac{\cos(m\cos^{-1}(\zeta))}{(\sin^{2}\theta\sin^{2}\theta'-(x-\cos\theta\cos\theta')^{2})^{1/2}}
\end{align}
where
\begin{equation}
\zeta=\frac{x-\cos\theta\cos\theta'}{\sin\theta\sin\theta'}
\end{equation}
and $x=\cos\lambda$ must lie in the range $\theta-\theta'<\lambda<\min\{\theta+\theta', 2\pi-\theta-\theta'\}$. For $\lambda$ outside of this range, the series sums to zero. An equivalent form of this result can be found in Ref.~\cite{Hansen} though our derivation generalizes the method of Baranov \cite{Baranov} who proves the $m=0$ case. This summation formula allows us to perform the $l$-sum in our Green's function expression yielding
\begin{widetext}
\begin{align}
\label{eq:greensfnlsum}
G_{\textrm{static}}(\textbf{x},\textbf{x}')=\frac{1}{4\pi^{2} b}\int_{x_{-}}^{x_{+}}\sum_{m=-\infty}^{\infty}\frac{\exp\left[-i m (\Delta\phi+\gamma\cosh^{-1}(\chi))\right]\cos\left[m  \cos^{-1}(\zeta)\right]\,d x}{(\sin^{2}\theta\sin^{2}\theta'-(\cos\theta\cos\theta'-x)^{2})^{1/2}((\eta\eta'-x)^{2}-(\eta^{2}-1)(\eta'^{2}-1))^{1/2}},
\end{align}
where the integral vanishes whenever $x$ lies outside the range $(x_{-}, x_{+})$ where $x_{\pm}=\cos\theta\cos\theta'\pm\sin\theta\sin\theta'$. The appearance of the integral may be greatly improved with the introduction of the independent variable $\Psi$, defined by
\begin{align}
x=\cos\lambda=\cos\theta\cos\theta'+\sin\theta\sin\theta'\cos\Psi,\quad \frac{d\Psi}{d x}=\pm\frac{1}{\sqrt{\sin^{2}\theta\sin^{2}\theta'-(x-\cos\theta\cos\theta')^{2}}}
\end{align}
where the choice of sign here is determined by the fact that $x$, as a function of $\Psi$, decreases from $x_{+}$ at the point $\Psi=0$ to $x_{-}$ at the point $\Psi=\pi$, and it increases symmetrically to $x_{+}$ at $\Psi=2\pi$. Therefore
\begin{align}
\label{eq:gstaticmsumintold}
G_{\textrm{static}}(\textbf{x},\textbf{x}')=&\frac{1}{16\pi^{2} b}\int_{0}^{2\pi}\sum_{m=-\infty}^{\infty}\frac{\exp\left(im(\Psi-\Delta\phi-\gamma\cosh^{-1}(\chi))\right)}{(\eta^{2}+\eta'^{2}-2\eta\eta'\cos\lambda-\sin^{2}\lambda)^{1/2}}d\Psi\nonumber\\
+&\frac{1}{16\pi^{2} b}\int_{0}^{2\pi}\sum_{m=-\infty}^{\infty}\frac{\exp\left(im(\Psi-\Delta\phi-\gamma\cosh^{-1}(\chi))\right)}{(\eta^{2}+\eta'^{2}-2\eta\eta'\cos\lambda-\sin^{2}\lambda)^{1/2}}d\Psi
\end{align}
where
\begin{equation}
\chi=\frac{\eta\eta'-\cos\lambda}{(\eta^{2}-1)^{1/2}(\eta'^{2}-1)^{1/2}}=\frac{\eta\eta'-\cos\theta\cos\theta'-\sin\theta\sin\theta'\cos\Psi}{(\eta^{2}-1)^{1/2}(\eta'^{2}-1)^{1/2}},
\end{equation}
and we have written the cosine in the numerator in Eq.(\ref{eq:greensfnlsum}) as a sum of exponentials. It is easy to see that these integrals are invariant under the change $\Psi\rightarrow-\Psi$ in the integrand so we obtain the single integral
\begin{equation}
\label{eq:gstaticmsumint}
G_{\textrm{static}}(\textbf{x},\textbf{x}')=\frac{1}{8\pi^{2} b}\int_{0}^{2\pi}\sum_{m=-\infty}^{\infty}\frac{\exp\left(im(\Psi-\Delta\phi-\gamma\cosh^{-1}(\chi))\right)}{(\eta^{2}+\eta'^{2}-2\eta\eta'\cos\lambda-\sin^{2}\lambda)^{1/2}}d\Psi.
\end{equation}
The $m=0$ term gives the static, axisymmetric Green's function
\begin{equation}
\label{eq:gkerrm0}
G_{\textrm{static}}^{m=0}(\eta,\theta,\eta',\theta')=\frac{1}{8\pi^{2} b}\int_{0}^{2\pi}\frac{1}{(\eta^{2}+\eta'^{2}-2\eta\eta'\cos\lambda-\sin^{2}\lambda)^{1/2}}d\Psi,
\end{equation}
which has been previously noted by Linet \cite{Linet1977} using axisymmetric potential theory. The $m$-sum can be performed using the Fourier representation of a periodic delta function
\begin{equation}
\delta(x)=\frac{1}{2\pi}\sum_{k=-\infty}^{\infty}e^{i k x},
\end{equation}
which gives
\begin{equation}
\label{eq:gstaticdeltaint}
G_{\textrm{static}}(\textbf{x}, \textbf{x}')=\frac{1}{4\pi b}\int_{0}^{2\pi}\frac{\delta(\Psi-\Delta\phi-\gamma\cosh^{-1}(\chi))}{(\eta^{2}+\eta'^{2}-2\eta\eta'\cos\lambda-\sin^{2}\lambda)^{1/2}}d\Psi.
\end{equation}
The integral is now somewhat trivial, since we pick up contributions only at the zeros of the delta function, though this is complicated by the fact that the zeros are solutions of a transcendental equation, which we cannot solve analytically. Nevertheless, the integral may be performed using the definition of the delta composition:
\begin{equation}
\label{eq:deltacomposition}
\delta(f(x))=\sum_{i} \frac{\delta(x-x_{i})}{|f'(x_{i})|}
\end{equation}
where the sum is over all roots $x_{i}$ of $f(x)=0$. In particular, we have
\begin{align}
\label{eq:deltacompositionpsi}
\delta(\Psi-\Delta\phi-\gamma\cosh^{-1}(\chi))=\sum_{i}\frac{\delta(\Psi-\Psi_{i})(\eta^{2}+\eta'^{2}-2\eta\eta'\cos\lambda_{i}-\sin^{2}\lambda_{i})^{1/2}}{|(\eta^{2}+\eta'^{2}-2\eta\eta'\cos\lambda_{i}-\sin^{2}\lambda_{i})^{1/2}-(a/b)\sin\theta\sin\theta'\sin\Psi_{i}|}
\end{align}
where $\cos\lambda_{i}=\cos\theta\cos\theta'+\sin\theta\sin\theta'\cos\Psi_{i}$ and $\Psi_{i}$ are solutions of the transcendental equation
\begin{equation}
\label{eq:psiroots}
\Psi_{i}=\Delta\phi+\gamma\cosh^{-1}\left(\frac{\eta\eta'-\cos\theta\cos\theta'-\sin\theta\sin\theta'\cos\Psi_{i}}{(\eta^{2}-1)^{1/2}(\eta'^{2}-1)^{1/2}}\right).
\end{equation}
The Green's function is therefore
\begin{align}
G_{\textrm{static}}(\textbf{x}, \textbf{x}')=\frac{1}{4\pi b}\sum_{i}\frac{1}{|(\eta^{2}+\eta'^{2}-2\eta\eta'\cos\lambda_{i}-\sin^{2}\lambda_{i})^{1/2}-(a/b)\sin\theta\sin\theta'\sin\Psi_{i}|}.
\end{align}
It turns out to be useful to re-write Eq.(\ref{eq:psiroots}) in the form
\begin{equation}
\label{eq:psirootsnew}
\cosh\Big(\frac{b}{a}(\Psi_{i}-\Delta\phi)\Big)=\frac{\eta\eta'-\cos\lambda_{i}}{(\eta^{2}-1)^{1/2}(\eta'^{2}-1)^{1/2}}
\end{equation}
from which we obtain
\begin{equation}
(\eta^{2}+\eta'^{2}-2\eta\eta'\cos\lambda_{i}-\sin^{2}\lambda_{i})^{1/2}=(\eta^{2}-1)^{1/2}(\eta'^{2}-1)^{1/2}\sinh\Big(\frac{b}{a}|\Psi_{i}-\Delta\phi |\Big)
\end{equation}
where we have assumed, without loss of generality, that $a>0$. This expression may now be used to greatly simplify our Green's function:
\begin{align}
\label{eq:greensfnsumroots}
G_{\textrm{static}}(\textbf{x}, \textbf{x}')=\frac{1}{4\pi b}\sum_{i}\frac{1}{|(\eta^{2}-1)^{1/2}(\eta'^{2}-1)^{1/2}\sinh(\tfrac{b}{a}|\Psi_{i}-\Delta\phi |)-(a/b)\sin\theta\sin\theta'\sin\Psi_{i}|}.
\end{align}
\end{widetext}
We now define
\begin{equation}
f(\Psi)=\cosh\Big(\frac{b}{a}(\Psi-\Delta\phi)\Big)-\frac{\eta\eta'-\cos\lambda}{(\eta^{2}-1)^{1/2}(\eta'^{2}-1)^{1/2}}
\end{equation}
so that the $\Psi_{i}$ are roots of $f(\Psi)$. Solving for the turning points, $f'(\Psi)=0$ is equivalent to solving
\begin{align}
\sinh\big(\tfrac{b}{a}(\Psi-\Delta\phi)\big)&=\frac{(a/b)\sin\theta\sin\theta'\sin\Psi}{(\eta^{2}-1)^{1/2}(\eta'^{2}-1)^{1/2}}\nonumber\\
&=\frac{b}{a}\frac{a \sin\theta}{\Delta^{1/2}}\frac{a \sin\theta'}{\Delta'^{1/2}}\sin\Psi.
\end{align}
For points outside the ergosphere, which are the only physically relevant points to consider for a static Green's function, we have the following inequalities
\begin{align}
\Delta^{1/2}>a \sin\theta,\quad\Delta'^{1/2}>a \sin\theta'\, ,\qquad \textrm{for}\,\,&a>0, \nonumber\\
0<&\theta<\pi,
\end{align}
so that the turning points $\Psi$ satisfies the inequality
\begin{equation}
-\frac{b}{a}<\sinh\big(\tfrac{b}{a}(\Psi-\Delta\phi)\big)<\frac{b}{a}.
\end{equation}
Since $\sinh x$ is monotonic, there is precisely one turning point in the interval
\begin{equation}
\Delta\phi-\tfrac{a}{b}\sinh^{-1}(\tfrac{b}{a})<\Psi<\Delta\phi+\tfrac{a}{b}\sinh^{-1}(\tfrac{b}{a}).
\end{equation}
Furthermore, it is easy to verify that this turning point is a minimum. Hence there are, in general, two solutions to the transcendental equation (\ref{eq:psiroots}) which are approximately symmetric about the turning point. It turns out, however, that one of these solutions ought to be discarded since its inclusion in the Green's function yields the wrong Hadamard singularity structure. We take the solution that lies to the right of the turning point (which means we can discard the absolute value sign enclosing the denominator of Eq.(\ref{eq:greensfnsumroots})) since only this root yields the correct Hadamard singularity structure for the retarded field. We can justify this mathematically by virtue of the ambiguity in the integral (\ref{eq:gstaticmsumint}) which may be taken over any interval of length $2\pi$ since the integrand is periodic; thus we simply choose any interval which includes the root of Eq.(\ref{eq:psiroots}) which gives the correct singularities but which excludes the other root. Taken over such an interval, the integral in Eq.(\ref{eq:gstaticdeltaint}) only picks up the contribution from the appropriate root. One can show that the discarded root actually corresponds to the advanced Green's function, which implies that the boundary conditions, which were completely fixed by the choice of Legendre functions in the mode-sum expression, became ambiguous again in the process of summation.

If we denote the appropriate root as $\Psi_{0}$, we obtain the following closed-form expression for the Green's function for a static source in the Kerr black hole space-time:
\begin{align}
\label{eq:greensfnstatickerr}
&G_{\textrm{static}}(\textbf{x}, \textbf{x}')\nonumber\\
&=\frac{b}{4\pi}\frac{1}{\Delta^{1/2}\Delta'^{1/2}\sinh(\tfrac{b}{a}|\Psi_{0}-\Delta\phi|)-ab\sin\theta\sin\theta'\sin\Psi_{0}}
\end{align}
where we have reverted back to Boyer-Lindquist coordinates.

\section{The Self-Force on a Static Scalar Charge in Kerr Space-Time}
In this section, we shall use our expression for the static Kerr Green's function obtained in the previous section to calculate the self-force on a static (with respect to an undragged, static observer at infinity) scalar particle in the Kerr background space-time. Since we have a closed-form expression for the appropriate Green's function, we can derive completely analytic expressions for the self-force. We restrict our attention to the simplest case of a point-like scalar charge $q$ of mass $m$ coupled to a massless scalar field $\Phi(x)$. This calculation has previously been considered in \cite{BurkoLiu} using the mode-sum regularization prescription (MSRP), developed by Barack, Ori and collaborators \cite{BarackOri, Barack2001, BarackMino, BarackSago}, as well as Detweiler and collaborators \cite{DetweilerWhiting, Detweiler, DetweilerVega}. The conjectured result for the self-force in \cite{BurkoLiu} is
\begin{equation}
\label{eq:burkoliu}
f_{b}^{\textrm{self}}=\frac{1}{3}q^{2}\frac{a\,M^{2}\Delta \sin^{2}\theta}{(\Delta-a^{2}\sin^{2}\theta)^{5/2}\Sigma^{1/2}}\delta^{\phi}{}_{b}.
\end{equation}
This derivation, however, is somewhat unsatisfactory in that the authors are forced to conjecture the unknown MSRP coefficient $D_{a}$ based on its known form in other space-times as well as inferring from numerical results that the contribution from the tail-integral is zero. By calculating the self-force by other more direct means, we can establish whether or not the authors' conjecture for the coefficient $D_{a}$ is correct (although this coefficient as well as higher order MSRP coefficients have been recently computed for the Schwarzschild \cite{Heffernan} and Kerr \cite{Heffernan2} space-times) and verify the accuracy of the numerical results. Moreover, since the MSRP appears to be an efficient method for calculating the self-force for more difficult trajectories such as circular and eccentric orbits in Kerr \cite{Barack2010, Barack2011}, it is important to establish the accuracy of this method by comparing with the results obtained by alternative methods. More recently, a method known as the $m$-mode regularization prescription has been developed and applied to the Kerr case \cite{BarackGolbourn, BarackDolan, DolanWardell} and again the calculation presented in this section may serve as a standard non-trivial check of the accuracy of the $m$-mode regularization scheme.

\subsection{General Considerations}
The massless scalar field, $\Phi(x)$, satisfies
\begin{equation}
\label{eq:fieldeqn}
(\Box-\xi R)\Phi(x)=-4\pi \rho(x)
\end{equation}
where $\Box$ is the wave-operator (d'Alembertian) on the background geometry, $R$ is the Ricci scalar which is zero for vacuum space-times such as Kerr and $\xi$ is the coupling of the field to the curvature. The charge density, $\rho(x)$, of the point-particle is
\begin{equation}
\rho(x)=q \int_{\gamma}\frac{\delta^{4}(x^{a}-z^{a}(\tau))}{\sqrt{g}}\,d\tau
\end{equation}
where $z(\tau)$ describes the worldline $\gamma$ of the particle with proper time $\tau$, $g=|\det(g_{ab})|$ where $g_{ab}$ are the background metric coefficients and $\delta^{4}(\cdot)$ is the four-dimensional Dirac distribution. The scalar field, $\Phi(x)$, moves on null geodesics of the background space-time whereas the scalar particle itself is massive and therefore moves along time-like curves of the background. In addition, the field may scatter off the curvature of the space-time and re-interact with the particle. This radiation reaction process gives rise to a self-force
\begin{equation}
f_{a}^{\textrm{self}}=q\nabla_{a}\Phi_{\textrm{R}}(z(\tau)).
\end{equation}
This self-force then appears on the right-hand side of the equations of motions for the scalar particle
\begin{equation}
m\,a^{b}=(g^{bc}+u^{b}u^{c})f_{c}^{\textrm{self}}=q(g^{bc}+u^{b}u^{c})\nabla_{c}\Phi_{\textrm{R}}(z(\tau))
\end{equation}
where $u^{b}=d z^{b}/d\tau$ and $a^{b}=Du^{b}/d\tau$ are the four-velocity and four-acceleration of the particle, respectively. The crucial step in obtaining the correct self-force is identifying the correct radiative field that is regular at the particle's position, which we have called $\Phi_{\textrm{R}}$. The mass, $m$, appearing in the equations of motion is the `dynamical' particle mass, which in the scalar case evolves according to
\begin{equation}
\frac{dm}{d\tau}=-q u^{b}\nabla_{b}\Phi_{\textrm{R}}(z(\tau)).
\end{equation}

An expression for the derivative of the radiative field that is regular at the particle's position can be obtained in terms of an integral of the retarded Green's function over the entire history of the particle's motion:
\begin{align}
\label{eq:radiativefieldgradient}
\nabla_{b}\Phi_{\textrm{R}}(z(\tau))=\Phi_{b}^{\textrm{tail}}(z(\tau))-\frac{1}{12}(1-6\xi)R\,q\,u_{b}\nonumber\\
+q(g_{bc}+u_{b}u_{c})\Big(\frac{1}{3}\dot{a}^{c}+\frac{1}{6}R^{c}{}_{d}u^{d}\Big)
\end{align}
where $R^{a}{}_{b}$ is the Ricci tensor of the background metric and $\dot{a}^{b}=\textrm{D} a^{b}/d\tau$ is the Fermi derivative with respect to proper time of the four-acceleration. The last two terms are purely geometrical and easily evaluated. The global radiative term, the so-called \textit{tail integral} term, $\Phi_{a}^{\textrm{tail}}$, is
\begin{equation}
\label{eq:tailintegral}
\Phi_{a}^{\textrm{tail}}(z(\tau))=q\lim_{\epsilon\rightarrow 0^{+}}\int_{-\infty}^{\tau-\epsilon}\nabla_{a}G_{\textrm{ret}}(z(\tau),z(\tau'))\,d\tau'
\end{equation}
where $G_{\textrm{ret}}(z,z')$ is the retarded scalar Green's function satisfying
\begin{equation}
(\Box-\xi R)G_{\textrm{ret}}(z,z')=-g^{-1/2}\delta^{4}(z-z').
\end{equation}

Since the Green's function is singular at coincidence $z=z'$, we require some limiting process in order to regularize the self-force. When $z$ is within a normal neighbourhood of $z'$, then the retarded Green's function has the following Hadamard representation \cite{PoissonLR}:
\begin{align}
G_{\textrm{ret}}(z,z')=\frac{1}{4\pi}\Big\{\Delta^{1/2}(z,z')\,\delta_{+}(\sigma(z,z'))\nonumber\\
-V(z,z')\,\Theta_{+}(-\sigma(z,z'))\Big\}
\end{align}
where $\sigma(z,z')$ is the Synge world function which is half the square of the geodesic distance between $z$ and $z'$, $\delta_{+}(\sigma)$ is the past light-cone delta distribution and $\Theta_{+}(-\sigma)$ is the past light-cone step function. The biscalar $\Delta(z,z')$ is the Van-Vleck Morrette determinant which is regular and symmetric, as is $V(z,z')$. The presence of the $\delta_{+}(\sigma(z,z'))$ in the direct part of the retarded Green's function implies that the direct part has support only on the past light-cone and therefore does not contribute to the self-force since the tail integral in Eq.(\ref{eq:tailintegral}) is entirely inside the past light-cone. Hence, it is the tail part of the Green's function that is responsible for the self-force.

\begin{figure}
\centering
\includegraphics[width=9cm]{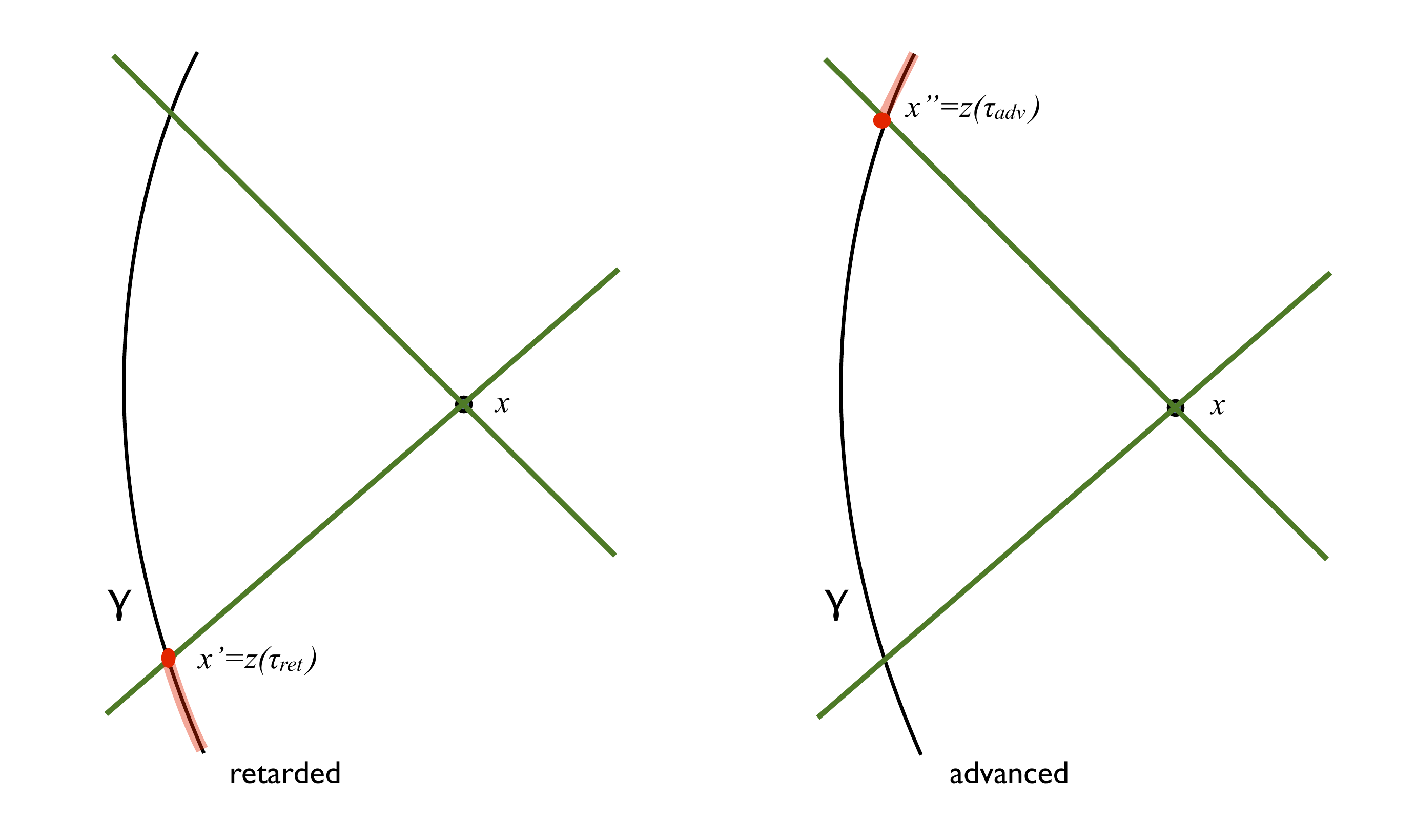}
\caption[The advanced and retarded field support.]{\emph{ The retarded field at $x$ depends on the entire particle's history from $-\infty<\tau\le\tau_{\textrm{ret}}$. The advanced field at $x$ depends on the entire future of the particle from $\tau_{\textrm{adv}}\le\tau<\infty$. The singular field depends only on the trajectory of the particle between the advanced and retarded points.}}
\label{fig:plot_Gret_Gadv}
\end{figure}

The causal regions are defined by the intersection of the past and future null cone of an arbitrary field point, $x$ say, with the world-line, $\gamma$, of the scalar charge. Following Poisson \cite{PoissonLR}, we shall call the intersection of the past light cone of $x$ with the world-line $\gamma$ the \textit{retarded point} associated with $x$ and we denote it $x'=z(\tau_{\textrm{ret}})$, where $\tau_{\textrm{ret}}$ is the \textit{retarded time}. Similarly, the future light cone intersects the world line at the \textit{advanced point} $x''=z(\tau_{\textrm{adv}})$, where $\tau_{\textrm{adv}}$ is the \textit{advanced time}. The retarded Green's function depends on the entire history of the particle up to the retarded point. One can define an advanced Green's function in an analogous way where the dependence is now on the entire future of the particle from the advanced point to infinity. This dependence is seen in Fig \ref{fig:plot_Gret_Gadv}.

\subsection{The Radiative Field for a Static Particle}
Rather than adopting the expression (\ref{eq:radiativefieldgradient}) for the gradient of the radiative field, in the static case it is more direct to begin with the definition of the radiative field as the difference between the retarded and singular fields:
\begin{align}
\label{eq:radiativefield}
\Phi_{\textrm{R}}(x)&=\Phi_{\textrm{ret}}(x)-\Phi_{\textrm{S}}(x)\nonumber\\
&=4\pi q\int_{\gamma}G_{\textrm{ret}}(x,z(\tau)) d\tau-4\pi q\int_{\gamma}G_{\textrm{S}}(x,z(\tau))d\tau,
\end{align}
where in order to calculate the self-force the radiative field must be evaluated on the world-line of the scalar particle. Detweiler and Whiting \cite{DetweilerWhiting2003} have identified the singular Green's function that yields the correct radiative field which has the following Hadamard construction:
\begin{equation}
\label{eq:Gs}
G_{\textrm{S}}(x,z)=\frac{1}{8\pi}\Big[\Delta^{1/2}(x,z)\delta(\sigma(x,z))+V(x,z)\Theta(\sigma(x,z))\Big].
\end{equation}
The singular (\textit{direct}) part of this Green's function has support only on the light cone due to the $\delta(\sigma)$ term. For $z(\tau)<z(\tau_{\textrm{ret}})$ and $z(\tau)>z(\tau_{\textrm{adv}})$, it is clear that $x$ and $z(\tau)$ are time-like related, meaning that $\sigma(x,z(\tau))<0$ in these regions. Moreover, for $\tau_{\textrm{ret}}<\tau<\tau_{\textrm{adv}}$, the points $x$ and $z(\tau)$ are space-like related and therefore $\sigma(x, z(\tau))>0$ in this region. These considerations combined with the presence of the $\Theta(\sigma)$ in the tail term of Eq.(\ref{eq:Gs}) imply that the singular Green's function has support only in the region $\tau_{\textrm{ret}}\le\tau\le\tau_{\textrm{adv}}$ (see Fig \ref{fig:plot_Gret_Gadv}).

We consider now the retarded field appearing in Eq.(\ref{eq:radiativefield}), we take the coordinates of the scalar charge at some proper-time $\tau$ to be $\bar{x}^{a}=z^{a}(\tau)$ and change the integration variable to run over the coordinate time $\bar{t}$ using $u^{\bar{t}}\,d\tau=d\bar{t}$, which gives
\begin{equation}
\Phi_{\textrm{ret}}(x)=4\pi q\int_{-\infty}^{\infty}G_{\textrm{ret}}(\Delta t; \textbf{x}, \bar{\textbf{x}})\frac{d\bar{t}}{u^{\bar{t}}},
\end{equation}
where $\Delta t=\bar{t}-t$,   $\textbf{x}$ is the spatial position of the field point and $\bar{\textbf{x}}$ is the spatial position of the scalar charge. Generally speaking, the spatial points will depend implicitly on the time, $\textbf{x}=\textbf{x}(t)$, but in the static case, by definition, there is no temporal dependence. For a stationary metric, the Green's function may be decomposed into Fourier frequency modes:
\begin{align}
\Phi_{\textrm{ret}}(x)=4\pi q\frac{1}{u^{\bar{t}}}\int_{-\infty}^{\infty}\int_{-\infty}^{\infty}e^{-i\omega\Delta t}G_{\omega}^{(3)}(\textbf{x},\bar{\textbf{x}})\,d\omega\, d\bar{t},
\end{align}
where $G_{\omega}^{(3)}$ is a three-dimensional Green's function on a dimensionally reduced metric. For a static particle, $G_{\omega}^{(3)}$ is independent of $\bar{t}$ and this integration may be performed, yielding a $\delta(\omega)$ term. The $\omega$ integral is now trivial yielding
\begin{equation}
\Phi_{\textrm{ret}}^{\textrm{static}}(x)=4\pi q\frac{1}{u^{\bar{t}}}G_{0}^{(3)}(\textbf{x},\bar{\textbf{x}})=4\pi q\frac{1}{u^{\bar{t}}}G_{\textrm{static}}(\textbf{x},\bar{\textbf{x}})
\end{equation}
where, in the particular case of the Kerr geometry, $G_{\textrm{static}}(\textbf{x},\bar{\textbf{x}})$ is given by Eq.(\ref{eq:greensfnstatickerr}).

Consider now the singular field. From Eq.(\ref{eq:radiativefield}) and Eq.(\ref{eq:Gs}), we obtain
\begin{align}
\Phi_{\textrm{S}}(x)=\frac{1}{2}q\int_{-\infty}^{\infty}\Big\{&\Delta^{1/2}(x,\bar{x})\delta(\sigma(x,\bar{x}))\nonumber\\
&-V(x,\bar{x})\Theta(\sigma(x,\bar{x}))\Big\}\,d\tau.
\end{align}
In the first integral here, we change the variable of integration to $\sigma$ using $\sigma_{;\bar{a}}u^{\bar{a}}d\tau=d\sigma$, where by $\sigma_{;\bar{a}}$ we mean covariant derivative with respect to the coordinates $\bar{x}^{a}$. The only contributions are when $\sigma=0$, i.e. at the advanced and retarded points. The second integral vanishes outside of the range enclosed by these two points as a consequence of the step function. Therefore, we have \cite{PoissonLR}
\begin{align}
&\Phi_{\textrm{S}}(x)\nonumber\\
&=q\Big(\frac{\Delta^{1/2}(x,x')}{2 \,r_{\textrm{ret}}}+\frac{\Delta^{1/2}(x,x'')}{2\, r_{\textrm{adv}}}+\frac{1}{2}\int_{\tau_{\textrm{ret}}}^{\tau_{\textrm{adv}}}V(x,\bar{x}(\tau)) d\tau\Big)
\end{align}
where, for a static particle, $r_{\textrm{ret}}=\sigma_{,t'}u^{t'}$ is the retarded distance between $x$ and the world-line and $r_{\textrm{adv}}=-\sigma_{,t''}u^{t''}$ is the advanced distance between $x$ and the world-line. The integral term is regular in the limit as $x$ approaches the world-line. The geometrical nature of the direct part is now explicit as the world-line is approached ($r\rightarrow 0$).

Therefore, for a static particle in a stationary space-time, the self-force is given by
\begin{align}
\label{eq:selfforcestatic}
f^{\textrm{self}}_{a}(\bar{x})=q^{2}\lim_{x\rightarrow \bar{x}}\Big[\nabla_{a}\Big(\frac{4\pi}{u^{\bar{t}}}G_{\omega=0}^{(3)}(\textbf{x}, \bar{\textbf{x}})- \frac{\Delta^{1/2}(x,x')}{2 \,r_{\textrm{ret}}}\nonumber\\
-\frac{\Delta^{1/2}(x,x'')}{2\, r_{\textrm{adv}}} -\frac{1}{2}\int_{\tau_{\textrm{ret}}}^{\tau_{\textrm{adv}}}V(x,\bar{x}(\tau)) d\tau\Big)\Big].
\end{align}
where $G_{\omega=0}^{(3)}$ is the zero frequency Fourier mode of the four-dimensional retarded Green's function. 

\subsection{Calculating the Retarded and Advanced Distance for a Static Particle}
In this Section, we shall outline our method for calculating the retarded and advanced distances, $r_{\textrm{ret}}$ and $r_{\textrm{adv}}$, respectively, for a static particle in a stationary space-time. Calculating these quantities requires evaluations of a derivative of the geodesic distance at the retarded point $x'$ and the advanced point $x''$, which are connected to the field point $x$ by a null geodesic defined by $\sigma(x,\bar{x})=0$. One may obtain a coordinate expansion of $\sigma$ by assuming the form
\begin{align}
\label{eq:sigmacoordinateexp}
2\sigma(x,\bar{x})= & g_{ab}\Delta x^{a}\Delta x^{b}+A_{abc} \Delta x^{a} \Delta x^{b} \Delta x^{c}\nonumber\\
&+B_{abcd}\Delta x^{a} \Delta x^{b} \Delta x^{c} \Delta x^{c} +\textrm{O}(\Delta x^{5})
\end{align}
and then substituting this \textit{ansatz} into the defining equation for $\sigma$
\begin{equation}
2 \sigma=g^{ab}\sigma_{;a}\sigma_{;b}.
\end{equation}
Equating powers of $\Delta x$ yields
\begin{align}
\label{eq:sigmatensors}
A_{abc}=\frac{1}{2}g_{(ab,c)},\quad B_{abcd}=\frac{1}{6}g_{(ab,cd)}-\frac{1}{12}g_{ef}\Gamma^{e}_{(ab}\Gamma^{f}_{cd)},
\end{align}
where, in our conventions $\Delta x^{a}=\bar{x}^{a}-x^{a}$, i.e., field point expanded about the particle world-line, and round brackets around indices imply symmetrization. Then
\begin{align}
\label{eq:sigmatbar}
2\sigma_{,\bar{t}}=& 2 g_{tt}\Delta t+2g_{\alpha t}\Delta x^{\alpha}+3 A_{ttt}\Delta t^{2}+6A_{\alpha tt}\Delta t \Delta x^{\alpha}\nonumber\\
&+3 A_{\alpha\beta t}\Delta x^{\alpha} \Delta x^{\beta}+4 B_{tttt}\Delta t^{3}+12 B_{\alpha ttt} \Delta x^{\alpha}\Delta t^{2}\nonumber\\
&+12 B_{\alpha\beta tt} \Delta x^{\alpha} \Delta x^{\beta}\Delta t+4 B_{\alpha\beta\gamma t}\Delta x^{\alpha} \Delta x^{\beta} \Delta x^{\gamma}\nonumber\\
&+\textrm{O}(\Delta x^{4})
\end{align}
which we wish to evaluate at the retarded and advanced points. This requires two additional steps: First, the tensors $g_{ab}$, $A_{abc}$ and $B_{abcd}$ are each evaluated at the field point $x$. However, we wish to evaluate them at the retarded and advanced points which are on the particle's world-line. Therefore, we Taylor expand these coefficients about the world-line. Second, we solve the equation $\sigma(x,\bar{x})=0$ iteratively to obtain the advanced and retarded times as expansions in $\Delta x^{\alpha}$. Substituting into our expressions for $\sigma_{,\bar{t}}$ gives us the advanced and retarded distances.

\begin{figure}
\centering
\includegraphics[width=8cm]{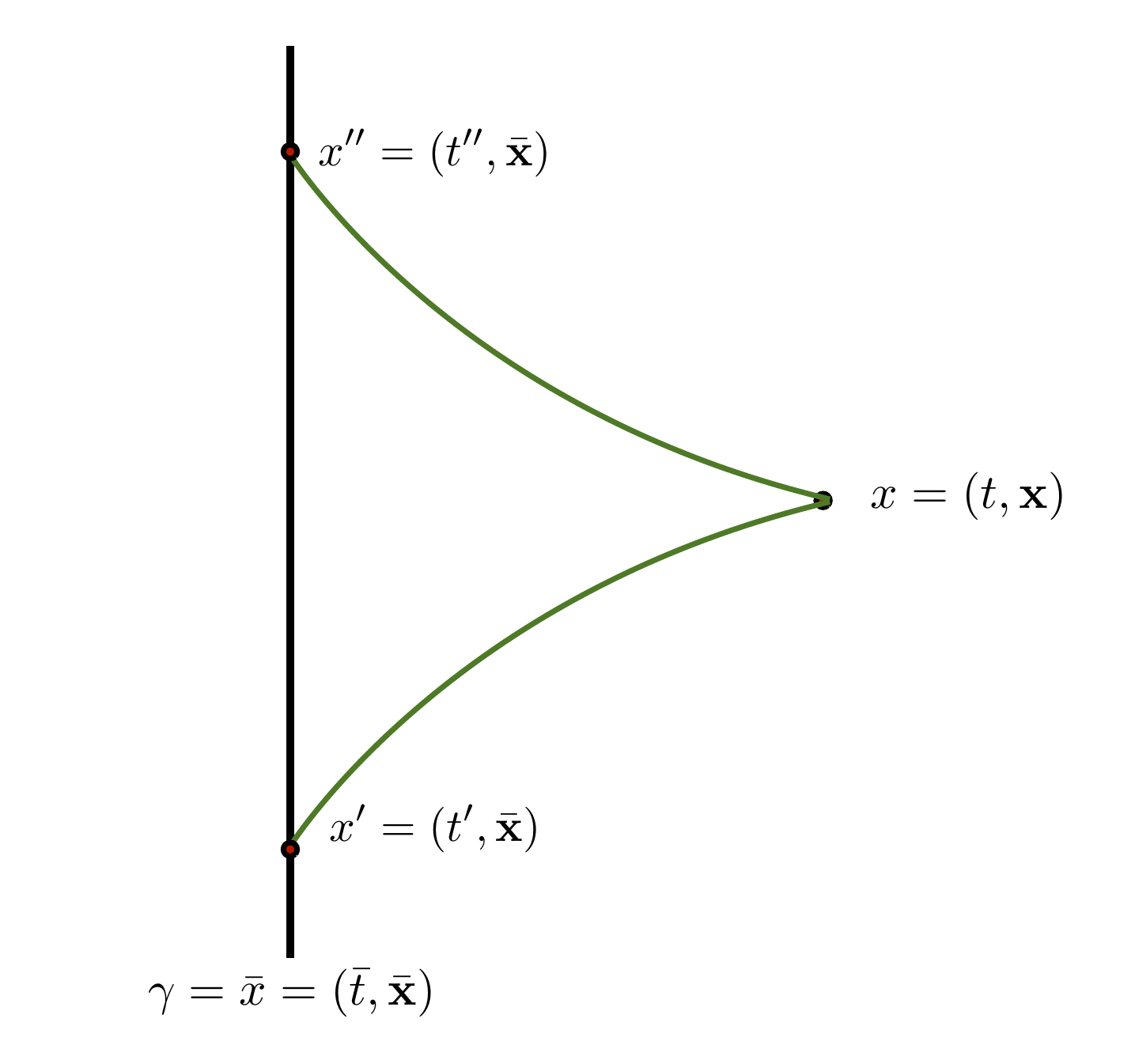}
\caption[The world-line of a static particle.]{\emph{The spatial coordinates of the world-line of a static particle are fixed over coordinate time.}}
\label{fig:plot_static_worldline}
\end{figure}

Taylor expanding the coefficients in expression (\ref{eq:sigmatbar}) about the world-line yields
\begin{align}
\label{eq:sigmatbartaylor}
2& \sigma_{,\bar{t}}=\epsilon(2 \bar{g}_{tt} \Delta t+2 \bar{g}_{\alpha t} \Delta x^{\alpha})+\epsilon^{2}(-2\bar{g}_{tt,\alpha}\Delta x^{\alpha} \Delta t\nonumber\\
&-2\bar{g}_{\alpha t,\beta}\Delta x^{\alpha}\Delta x^{\beta}+3 \bar{A}_{ttt}\Delta t^{2}+6 \bar{A}_{\alpha tt}\Delta t \Delta x^{\alpha}\nonumber\\
&+3 \bar{A}_{\alpha\beta t}\Delta x^{\alpha} \Delta x^{\beta})+\epsilon^{3}(\bar{g}_{tt,\alpha\beta}\Delta t\Delta x^{\alpha} \Delta x^{\beta}\nonumber\\
&+\bar{g}_{\alpha t,\beta\gamma}\Delta x^{\alpha} \Delta x^{\beta} \Delta x^{\gamma}-3 \bar{A}_{ttt,\alpha}\Delta x^{\alpha}\Delta t^{2}\nonumber\\
&-6 \bar{A}_{\alpha tt,\beta}\Delta t \Delta x^{\alpha} \Delta x^{\beta}-3 \bar{A}_{\alpha\beta t,\gamma}\Delta x^{\alpha}\Delta x^{\beta} \Delta x^{\gamma}\nonumber\\
&+4\bar{B}_{tttt}\Delta t^{3}+12\bar{B}_{\alpha ttt}\Delta t^{2} \Delta x^{\alpha}\nonumber\\
&+12 \bar{B}_{\alpha\beta tt} \Delta x^{\alpha} \Delta x^{\beta} \Delta t+4\bar{B}_{\alpha\beta\gamma t}\Delta x^{\alpha}\Delta x^{\beta}\Delta x^{\gamma})+\textrm{O}(\epsilon^{4}),
\end{align}
where an overbar means that the tensor is evaluated at $\bar{x}$ which is on the world-line and we have introduced a book-keeping parameter, $\epsilon$, in order to keep account of powers of $\Delta x$. It is worth noting that there is no need to distinguish metric dependent tensors $\bar{g}_{ab}$, $\bar{A}_{abc}$ and $\bar{B}_{abcd}$ from those evaluated at $x'$ or $x''$ since these tensors depend only on the spatial coordinates for any metrics of interest, which have the same spatial dependence for a static particle (see Fig \ref{fig:plot_static_worldline}). The difference in evaluating Eq.(\ref{eq:sigmatbartaylor}) at various points on the world-line of a static particle arises in the $\Delta t$ term. 

The difference $\Delta t$ corresponding to the advanced and retarded times we shall denote $\Delta t_{\textrm{adv}}$ and $\Delta t_{\textrm{ret}}$, respectively. These times are defined by the intersection of the null light cones with the world-line $\gamma$ and we are therefore required to solve $\sigma(x,\bar{x})=0$ as a function of $\Delta t$. A coordinate expansion for $\sigma$ has been given in Eq.(\ref{eq:sigmacoordinateexp}) along with expressions (\ref{eq:sigmatensors}). As before, the metric dependent tensors $A_{abc}$ and $B_{abcd}$ ought to be evaluated on the world-line and so a Taylor expansion about the world-line is required, yielding
\begin{align}
\label{eq:sigmataylorexp}
2& \sigma(x,\bar{x})=\epsilon^{2}(\bar{g}_{tt}\Delta t^{2}+2 \bar{g}_{t\alpha} \Delta t \Delta x^{\alpha}+\bar{g}_{\alpha\beta}\Delta x^{\alpha} \Delta x^{\beta})\nonumber\\
&+\epsilon^{3}(\bar{A}_{ttt}\Delta t^{3}+[3 \bar{A}_{tt\alpha}-\bar{g}_{tt,\alpha}]\Delta t^{2} \Delta x^{\alpha}\nonumber\\
&+[3\bar{A}_{t\alpha\beta}-2\bar{g}_{t\alpha ,\beta}]\Delta t \,\Delta x^{\alpha} \Delta x^{\beta}\nonumber\\
&+[\bar{A}_{\alpha\beta\gamma}-\bar{g}_{\alpha\beta,\gamma}]\Delta x^{\alpha}\Delta x^{\beta}\Delta x^{\gamma})\nonumber\\
&+\epsilon^{4}(\bar{B}_{tttt}\Delta t^{4}+[4\bar{B}_{ttt\alpha}-\bar{A}_{ttt,\alpha}]\Delta t^{3}\Delta x^{\alpha}\nonumber\\
&+[6\bar{B}_{tt\alpha\beta}-3\bar{A}_{tt\alpha,\beta}+\tfrac{1}{2}\bar{g}_{tt,\alpha\beta}]\Delta t^{2}\Delta x^{\alpha}\Delta x^{\beta}\nonumber\\
&+[4\bar{B}_{t\alpha\beta\gamma}-3\bar{A}_{t\alpha\beta,\gamma}+\bar{g}_{t\alpha,\beta\gamma}]\Delta t\,\Delta x^{\alpha}\Delta x^{\beta}\Delta x^{\gamma} \nonumber\\
&+[\bar{B}_{\alpha\beta\gamma\delta}-\bar{A}_{\alpha\beta\gamma ,\delta}+\tfrac{1}{2}\bar{g}_{\alpha\beta ,\gamma\delta}]\Delta x^{\alpha}\Delta x^{\beta}\Delta x^{\gamma}\Delta x^{\delta}) +\textrm{O}(\epsilon^{5}).
\end{align}
We wish to solve $\sigma=0$ iteratively for $\Delta t$ so we further expand
\begin{equation}
\label{eq:deltatexp}
\Delta t=\Delta t_{1}+\epsilon \Delta t_{2}+\epsilon^{2} \Delta t_{3}.
\end{equation}
Upon substitution of Eq.(\ref{eq:deltatexp}) into (\ref{eq:sigmataylorexp}), we obtain
\begin{align}
2&\sigma(x,\bar{x})=\epsilon^{2}(\bar{g}_{tt}\Delta t_{1}^{2}+2 \bar{g}_{t\alpha} \Delta t_{1} \Delta x^{\alpha}+\bar{g}_{\alpha\beta}\Delta x^{\alpha} \Delta x^{\beta})\nonumber\\
&+\epsilon^{3}(\bar{A}_{ttt}\Delta t_{1}^{3}+[3 \bar{A}_{tt\alpha}-\bar{g}_{tt,\alpha}]\Delta t_{1}^{2} \Delta x^{\alpha}\nonumber\\
&+[3\bar{A}_{t\alpha\beta}-2\bar{g}_{t\alpha ,\beta}]\Delta t_{1} \,\Delta x^{\alpha} \Delta x^{\beta}\nonumber\\
&+[\bar{A}_{\alpha\beta\gamma}-\bar{g}_{\alpha\beta,\gamma}]\Delta x^{\alpha}\Delta x^{\beta}\Delta x^{\gamma}+2\bar{g}_{tt}\Delta t_{1}\,\Delta t_{2}\nonumber\\
&+2\bar{g}_{t\alpha}\Delta t_{2}\,\Delta x^{\alpha})+\epsilon^{4}(\bar{B}_{tttt}\Delta t_{1}^{4}\nonumber\\
&+[4\bar{B}_{ttt\alpha}-\bar{A}_{ttt,\alpha}]\Delta t_{1}^{3}\Delta x^{\alpha}\nonumber\\
&+[6\bar{B}_{tt\alpha\beta}-3\bar{A}_{tt\alpha,\beta}+\tfrac{1}{2}\bar{g}_{tt,\alpha\beta}]\Delta t_{1}^{2}\Delta x^{\alpha}\Delta x^{\beta}\nonumber\\
&+[4\bar{B}_{t\alpha\beta\gamma}-3\bar{A}_{t\alpha\beta,\gamma}+\bar{g}_{t\alpha,\beta\gamma}]\Delta t_{1}\,\Delta x^{\alpha}\Delta x^{\beta}\Delta x^{\gamma} \nonumber\\
&+[\bar{B}_{\alpha\beta\gamma\delta}-\bar{A}_{\alpha\beta\gamma ,\delta}+\tfrac{1}{2}\bar{g}_{\alpha\beta ,\gamma\delta}]\Delta x^{\alpha}\Delta x^{\beta}\Delta x^{\gamma}\Delta x^{\delta}\nonumber\\
&+3\bar{A}_{ttt}\Delta t_{1}^{2}\Delta t_{2}+[6\bar{A}_{tt\alpha}-2\bar{g}_{tt,\alpha}]\Delta t_{1}\Delta t_{2}\Delta x^{\alpha}\nonumber\\
&+[3\bar{A}_{t\alpha\beta}-2\bar{g}_{t\alpha ,\beta}]\Delta t_{2}\Delta x^{\alpha}\Delta x^{\beta}+2\bar{g}_{tt}\Delta t_{1}\Delta t_{3}\nonumber\\
&+\bar{g}_{tt}\Delta t_{2}^{2}+2\bar{g}_{t\alpha}\Delta t_{3}\Delta x^{\alpha}) +\textrm{O}(\epsilon^{5}).
\end{align}
Setting $\sigma(x, \bar{x})$ equal to zero and equating equal orders of $\epsilon$ gives an iterative solution for $\Delta t$. In particular, at $\textrm{O}(\epsilon^{2})$, we have
\begin{equation}
\bar{g}_{tt}\Delta t_{1}^{2}+2 \bar{g}_{t\alpha} \Delta t_{1} \Delta x^{\alpha}+\bar{g}_{\alpha\beta}\Delta x^{\alpha} \Delta x^{\beta}=0,
\end{equation}
which is a quadratic in $\Delta t_{1}$ with solutions
\begin{equation}
\label{eq:deltatfirstorder}
\Delta t_{1}^{\pm}=\frac{-\bar{g}_{t\alpha}\Delta x^{\alpha}\pm\sqrt{(\bar{g}_{t\beta}\Delta x^{\beta})^{2}-\bar{g}_{tt}\bar{g}_{\alpha\beta}\Delta x^{\alpha}\Delta x^{\beta}}}{\bar{g}_{tt}}.
\end{equation}
These two solutions correspond to the first order approximation to the advanced and retarded times. From Fig \ref{fig:plot_static_worldline}, it is clear that $\Delta t_{\textrm{ret}}=(t'-t)<0$ and $\Delta t_{\textrm{adv}}=(t''-t)>0$. Since $\bar{g}_{ta}<0$ ($a=0...3$), the appropriate identification of the retarded and advanced times is
\begin{align}
\label{eq:deltat1}
\Delta t_{1, \textrm{ret}}&=\Delta t_{1}^{+} \nonumber\\
\Delta t_{1, \textrm{adv}}&=\Delta t_{1}^{-}.
\end{align}
The higher order corrections are linear in the dependent variables and so it is straight forward to calculate the $\textrm{O}(\epsilon^{3})$ and $\textrm{O}(\epsilon^{4})$ solutions:
\begin{align}
\label{eq:deltat21}
\Delta& t_{2,\textrm{ret}}=-\frac{1}{2(\bar{g}_{tt} \Delta t_{1,\textrm{ret}}+\bar{g}_{t\alpha}\Delta x^{\alpha})}\Big( \bar{A}_{ttt}\Delta t_{1,\textrm{ret}}^{3}\nonumber\\
&+[3 \bar{A}_{tt\alpha}-\bar{g}_{tt,\alpha}]\Delta t_{1, \textrm{ret}}^{2} \Delta x^{\alpha}\nonumber\\
&+[3\bar{A}_{t\alpha\beta}-2\bar{g}_{t\alpha ,\beta}]\Delta t_{1, \textrm{ret}} \,\Delta x^{\alpha} \Delta x^{\beta}\nonumber\\
&+[\bar{A}_{\alpha\beta\gamma}-\bar{g}_{\alpha\beta,\gamma}]\Delta x^{\alpha}\Delta x^{\beta}\Delta x^{\gamma}\Big) \\
\label{eq:deltat22}
\Delta& t_{2,\textrm{adv}}=-\frac{1}{2(\bar{g}_{tt} \Delta t_{1,\textrm{adv}}+\bar{g}_{t\alpha}\Delta x^{\alpha})}\Big( \bar{A}_{ttt}\Delta t_{1,\textrm{adv}}^{3}\nonumber\\
&+[3 \bar{A}_{tt\alpha}-\bar{g}_{tt,\alpha}]\Delta t_{1, \textrm{adv}}^{2} \Delta x^{\alpha}\nonumber\\
&+[3\bar{A}_{t\alpha\beta}-2\bar{g}_{t\alpha ,\beta}]\Delta t_{1, \textrm{adv}} \,\Delta x^{\alpha} \Delta x^{\beta}\nonumber\\
&+[\bar{A}_{\alpha\beta\gamma}-\bar{g}_{\alpha\beta,\gamma}]\Delta x^{\alpha}\Delta x^{\beta}\Delta x^{\gamma}\Big) \\
\label{eq:deltat31}
\Delta& t_{3, \textrm{ret}}=-\frac{1}{2(\bar{g}_{tt} \Delta t_{1, \textrm{ret}}+\bar{g}_{t\alpha}\Delta x^{\alpha})}\Big(\bar{B}_{tttt}\Delta t_{1, \textrm{ret}}^{4}\nonumber\\
&+[4\bar{B}_{ttt\alpha}-\bar{A}_{ttt,\alpha}]\Delta t_{1, \textrm{ret}}^{3}\Delta x^{\alpha}\nonumber\\
&+[6\bar{B}_{tt\alpha\beta}-3\bar{A}_{tt\alpha,\beta}+\tfrac{1}{2}\bar{g}_{tt,\alpha\beta}]\Delta t_{1, \textrm{ret}}^{2}\Delta x^{\alpha}\Delta x^{\beta} \nonumber\\
&+[4\bar{B}_{t\alpha\beta\gamma}-3\bar{A}_{t\alpha\beta,\gamma}+\bar{g}_{t\alpha,\beta\gamma}]\Delta t_{1, \textrm{ret}}\,\Delta x^{\alpha}\Delta x^{\beta}\Delta x^{\gamma}  \nonumber\\
&+[\bar{B}_{\alpha\beta\gamma\delta}-\bar{A}_{\alpha\beta\gamma ,\delta}+\tfrac{1}{2}\bar{g}_{\alpha\beta ,\gamma\delta}]\Delta x^{\alpha}\Delta x^{\beta}\Delta x^{\gamma}\Delta x^{\delta}\nonumber\\
&+[6\bar{A}_{tt\alpha}-2\bar{g}_{tt,\alpha}]\Delta t_{1, \textrm{ret}}\Delta t_{2, \textrm{ret}}\Delta x^{\alpha}\nonumber\\
&+[3\bar{A}_{t\alpha\beta}-2\bar{g}_{t\alpha ,\beta}]\Delta t_{2, \textrm{ret}}\Delta x^{\alpha}\Delta x^{\beta}\nonumber\\
&+3\bar{A}_{ttt}\Delta t_{1, \textrm{ret}}^{2}\Delta t_{2, \textrm{ret}}+\bar{g}_{tt}\Delta t_{2, \textrm{ret}}^{2}\Big)\\
\label{eq:deltat32}
\Delta& t_{3, \textrm{adv}}=-\frac{1}{2(\bar{g}_{tt} \Delta t_{1, \textrm{adv}}+\bar{g}_{t\alpha}\Delta x^{\alpha})}\Big(\bar{B}_{tttt}\Delta t_{1, \textrm{adv}}^{4}\nonumber\\
&+[4\bar{B}_{ttt\alpha}-\bar{A}_{ttt,\alpha}]\Delta t_{1, \textrm{adv}}^{3}\Delta x^{\alpha}\nonumber\\
&+[6\bar{B}_{tt\alpha\beta}-3\bar{A}_{tt\alpha,\beta}+\tfrac{1}{2}\bar{g}_{tt,\alpha\beta}]\Delta t_{1, \textrm{adv}}^{2}\Delta x^{\alpha}\Delta x^{\beta} \nonumber\\
&+[4\bar{B}_{t\alpha\beta\gamma}-3\bar{A}_{t\alpha\beta,\gamma}+\bar{g}_{t\alpha,\beta\gamma}]\Delta t_{1, \textrm{adv}}\,\Delta x^{\alpha}\Delta x^{\beta}\Delta x^{\gamma}  \nonumber\\
&+[\bar{B}_{\alpha\beta\gamma\delta}-\bar{A}_{\alpha\beta\gamma ,\delta}+\tfrac{1}{2}\bar{g}_{\alpha\beta ,\gamma\delta}]\Delta x^{\alpha}\Delta x^{\beta}\Delta x^{\gamma}\Delta x^{\delta}\nonumber\\
&+[6\bar{A}_{tt\alpha}-2\bar{g}_{tt,\alpha}]\Delta t_{1, \textrm{adv}}\Delta t_{2, \textrm{adv}}\Delta x^{\alpha}\nonumber\\
&+[3\bar{A}_{t\alpha\beta}-2\bar{g}_{t\alpha ,\beta}]\Delta t_{2, \textrm{adv}}\Delta x^{\alpha}\Delta x^{\beta}\nonumber\\
&+3\bar{A}_{ttt}\Delta t_{1, \textrm{adv}}^{2}\Delta t_{2, \textrm{ret}}+\bar{g}_{tt}\Delta t_{2, \textrm{adv}}^{2}\Big).
\end{align}

We are now in a position to write down expressions for the retarded and advanced distances as coordinate expansions. Combining Eq.(\ref{eq:sigmatbartaylor}) and Eq.(\ref{eq:deltatexp}) with Eqs.(\ref{eq:deltatfirstorder})-(\ref{eq:deltat32}) implies
\begin{align}
\label{eq:retardeddistance}
r_{\textrm{ret}}=\sigma_{,t'}u^{t'}=\frac{1}{2\sqrt{-\bar{g}_{tt}}}\Big(P_{\textrm{ret}}\epsilon+Q_{\textrm{ret}}\epsilon^{2}+R_{\textrm{ret}}\epsilon^{3}\Big)
\end{align}
where
\begin{align}
\label{eq:Pret}
&P_{\textrm{ret}}=2 \bar{g}_{tt} \Delta t_{1,\textrm{ret}}+2\bar{g}_{t\alpha}\Delta x^{\alpha},\\
\label{eq:Qret}
&Q_{\textrm{ret}}=3 \bar{A}_{ttt}\Delta t_{1,\textrm{ret}}^{2}+[6\bar{A}_{tt\alpha}-2\bar{g}_{tt,\alpha}]\Delta t_{1, \textrm{ret}} \Delta x^{\alpha}\nonumber\\
&\quad+[3\bar{A}_{t\alpha\beta}-2\bar{g}_{t\alpha , \beta}]\Delta x^{\alpha} \Delta x^{\beta}+2\bar{g}_{tt}\Delta t_{2, \textrm{ret}},
\end{align}
\begin{align}
\label{eq:Rret}
&R_{\textrm{ret}}=4\bar{B}_{tttt}\Delta t_{1, \textrm{ret}}^{3}+[12\bar{B}_{ttt\alpha}-3\bar{A}_{ttt,\alpha}]\Delta t_{1,\textrm{ret}}^{2}\Delta x^{\alpha}\nonumber\\
&\quad+[12\bar{B}_{tt\alpha\beta}-6\bar{A}_{tt\alpha,\beta}+\bar{g}_{tt,\alpha\beta}]\Delta t_{1, \textrm{ret}} \Delta x^{\alpha} \Delta x^{\beta}\nonumber\\
&\quad+[4\bar{B}_{t\alpha\beta\gamma}-3\bar{A}_{t\alpha\beta,\gamma}+\bar{g}_{t\alpha,\beta\gamma}]\Delta x^{\alpha}\Delta x^{\beta} \Delta x^{\gamma}\nonumber\\
&\quad+6\bar{A}_{ttt}\Delta t_{1, \textrm{ret}} \Delta t_{2, \textrm{ret}}+[6\bar{A}_{tt\alpha}-2\bar{g}_{tt,\alpha}]\Delta t_{2, \textrm{ret}} \Delta x^{\alpha}\nonumber\\
&\quad+2\bar{g}_{tt}\Delta t_{3, \textrm{ret}}.
\end{align}
Analogously
\begin{align}
\label{eq:advanceddistance}
r_{\textrm{adv}}=-\sigma_{,t''}u^{t''}=-\frac{1}{2\sqrt{-\bar{g}_{tt}}}\Big(P_{\textrm{adv}}\epsilon+Q_{\textrm{adv}}\epsilon^{2}+R_{\textrm{adv}}\epsilon^{3}\Big)
\end{align}
where $P_{\textrm{adv}}$, $Q_{\textrm{adv}}$ and $R_{\textrm{adv}}$ are obtained from Eqs.(\ref{eq:Pret})-(\ref{eq:Rret}) by replacing the $\Delta t_{\textrm{ret}}$ expressions with $\Delta t_{\textrm{adv}}$ terms.

We may now use these expressions for the advanced and retarded distances in Eq.(\ref{eq:selfforcestatic}). We first note that covariant expansions for the bitensors $\Delta^{1/2}(x,\bar{x})$ and $V(x,\bar{x})$ are found in \cite{Decanini:2008}, which for a massless field in a Ricci-flat space-time are given by
\begin{align}
\Delta^{1/2}(x,\bar{x})&=1+\textrm{O}(\Delta x^{4}) \nonumber\\
V(x,\bar{x})&=\textrm{O}(\Delta x^{2}),
\end{align}
which implies that the integral term involving the bitensor $V(x,\bar{x})$ in Eq.(\ref{eq:selfforcestatic}) will vanish in the limit $x\rightarrow \bar{x}$ and so will not contribute to the self-force. The self-force reduces to
\begin{equation}
\label{eq:selfforcestaticvacuum}
f^{\textrm{self}}_{a}=q^{2}\lim_{x\rightarrow \bar{x}}\left[\nabla_{a}\left(\frac{4\pi}{u^{\bar{t}}}G_{\omega=0}^{(3)}(\textbf{x}, \bar{\textbf{x}})- \frac{1}{2 \,r_{\textrm{ret}}}-\frac{1}{2\, r_{\textrm{adv}}}\right)\right].
\end{equation}
Substituting our expansions for the retarded and advanced distances in Eq.(\ref{eq:retardeddistance}) and Eq.(\ref{eq:advanceddistance}), respectively, we obtain the self-force on a static, massless scalar charge in a stationary, Ricci-flat space-time:
\begin{align}
\label{eq:selfforcestaticvacuumexp}
f^{\textrm{self}}_{a}=4\pi q^{2}\lim_{x\rightarrow \bar{x}}\bigg[ \sqrt{-\bar{g}_{tt}}\,\,\nabla_{a}\bigg(G_{\omega=0}^{(3)}(\textbf{x}, \bar{\textbf{x}})- G_{\textrm{sing}}(\textbf{x}, \bar{\textbf{x}})\bigg)\bigg]
\end{align}
where
\begin{align}
\label{eq:GsingPQR}
G_{\textrm{sing}}(\textbf{x}, \bar{\textbf{x}})&=\frac{1}{4\pi}\Big[\frac{1}{P_{\textrm{ret}}}-\frac{1}{P_{\textrm{adv}}}-\frac{Q_{\textrm{ret}}}{P_{\textrm{ret}}^{2}}+\frac{Q_{\textrm{adv}}}{P_{\textrm{adv}}^{2}} \nonumber\\
&+\frac{(Q_{\textrm{ret}}^{2}-P_{\textrm{ret}}R_{\textrm{ret}})}{P_{\textrm{ret}}^{3}}-\frac{(Q_{\textrm{adv}}^{2}-P_{\textrm{adv}}R_{\textrm{adv}})}{P_{\textrm{adv}}^{3}}\Big].
\end{align}

It is easy to check that Eq.(\ref{eq:selfforcestaticvacuumexp})-(\ref{eq:GsingPQR}) gives us zero self-force for a static particle in Schwarzschild, as expected \cite{Wiseman, Rosenthal}.

\subsection{Self-Force on a Static Particle in the Kerr Space-Time}
We now calculate the self-force on a scalar charge with arbitrary fixed spatial Boyer Lindquist coordinates outside the ergosphere (the notion of a static particle breaks down on the ergosphere where the Killing vector $\partial/\partial t$ becomes null). The algebra involved is unnecessarily cumbersome for arbitrary separations, so we calculate each component of the self-force separately, which allows us to set two of the three separations to zero. For example, if we wish to calculate the radial component of the self-force, then from Eq.(\ref{eq:selfforcestaticvacuumexp}) we require partial radial derivatives of the Green's function and the singular field before taking a coincidence limit. However, in this case, one can take partial coincidence limits $\Delta \phi=0$ and $\Delta \theta=0$ from the outset since these terms just pass through radial derivatives. This procedure makes the calculation more palatable from an algebraic perspective.

There are several steps involved in the calculation: First, we derive analytic expressions for $\Psi_{0}$ for small separations. We then use our results for $\Psi_{0}$ to obtain analytic expressions for the Green's function for the various separations. Then the singular terms are calculated. Finally we compute derivatives of the regular field and take coincidence limits to obtain the self-force.

\subsubsection{Calculating $\Psi_{0}$ for Small Separations}
In order to calculate the Green's function up to order $\textrm{O}(\Delta x)$ (which is what is required for the self-force), we require $\Psi_{0}$ up to $\textrm{O}(\Delta x^{3})$, where $\Psi_{0}$ satisfies the transcendental equation (\ref{eq:psirootsnew}), which in Boyer Lindquist coordinates is
\begin{equation}
\label{eq:psizerobl}
\cosh\Big(\frac{b}{a}(\Psi_{0}-\Delta\phi)\Big)=\frac{(r-M)(\bar{r}-M)-b^{2}\cos\lambda}{\Delta^{1/2}\bar{\Delta}^{1/2}},
\end{equation}
where
\begin{equation}
\cos\lambda=\cos\theta\cos\bar{\theta}+\sin\theta\sin\bar{\theta}\cos\Psi_{0}.
\end{equation}
As discussed in the previous section, we cannot solve this equation analytically in general. However, we shall show that it can be solved for small separations up to any order. Reintroducing the book-keeping parameter $\epsilon$ to keep track of orders of $\Delta x$, we write
\begin{equation}
\label{eq:psizeroexp}
\Psi_{0}=\Psi_{0}^{(1)} \epsilon+ \Psi_{0}^{(2)}\epsilon^{2}+\Psi_{0}^{(3)}\epsilon^{3}+\textrm{O}(\epsilon^{4}).
\end{equation}
Substituting this into Eq.(\ref{eq:psizerobl}), writing $r=\bar{r}-\epsilon \Delta r$, $\theta=\bar{\theta}-\epsilon \Delta \theta$, $\phi=\bar{\phi}-\epsilon \Delta \phi$, expanding about $\epsilon=0$ and then equating equal orders of $\epsilon$ gives a simple set of recursion relations for $\Psi_{0}^{(1)}$, $\Psi_{0}^{(2)}$ etc. For example, equating powers of $\epsilon^{2}$ yields the following quadratic in $\Psi_{0}^{(1)}$:
\begin{align}
(\Psi_{0}^{(1)})^{2}-\frac{2\bar{\Delta}\,\Psi_{0}^{(1)}\Delta\phi}{\bar{\Delta}-a^{2}\sin^{2}\bar{\theta}}+\frac{(\bar{\Delta}^{2}\,\Delta\phi^{2}-a^{2}s^{2})}{\bar{\Delta}(\bar{\Delta}-a^{2}\sin^{2}\bar{\theta})}=0
\end{align}
where
\begin{equation}
s^{2}=\Delta r^{2}+\Delta\theta^{2}\,\bar{\Delta}.
\end{equation}
The two solutions of this quadratic are
\begin{align}
&\Psi_{0, \pm}^{(1)}=\nonumber\\
&\frac{\bar{\Delta}\,\Delta\phi}{\bar{\Delta}-a^{2}\sin^{2}\bar{\theta}}\pm\sqrt{\frac{a^{2}\bar{\Delta}\sin^{2}\bar{\theta}\,\Delta\phi^{2}}{(\bar{\Delta}-a^{2}\sin^{2}\bar{\theta})^{2}}+\frac{a^{2}s^{2}}{\bar{\Delta}(\bar{\Delta}-a^{2}\sin^{2}\bar{\theta})}},
\end{align}
where we recall that $\Psi_{0}$ is the particular choice that yields the correct singular field, and as explained in Sec.~\ref{sec:closedformgreensfn} corresponds to choosing the more positive solution:
\begin{align}
&\Psi_{0}^{(1)}=\nonumber\\
&\frac{\bar{\Delta}\,\Delta\phi}{\bar{\Delta}-a^{2}\sin^{2}\bar{\theta}}+\sqrt{\frac{a^{2}\bar{\Delta}\sin^{2}\bar{\theta}\,\Delta\phi^{2}}{(\bar{\Delta}-a^{2}\sin^{2}\bar{\theta})^{2}}+\frac{a^{2}s^{2}}{\bar{\Delta}(\bar{\Delta}-a^{2}\sin^{2}\bar{\theta})}}.
\end{align}

The higher order coefficients are linear in the dependent variable and therefore uniquely defined in terms of the appropriate $\Psi_{0}^{(1)}$ solution. However, for arbitrary separations they become increasingly cumbersome from an algebraic perspective so we calculate each component of the self-force individually by specializing to a particular separation.

Throughout the remainder of this section we shall assume, without loss of generality, that $\Delta r>0$, $\Delta\theta>0$, $\Delta\phi>0$ and $a>0$. Then the limit in Eq.(\ref{eq:selfforcestaticvacuumexp}) is a one-sided limit from the left. In fact, the two-sided limit does not exist since the Green's function expansion for small separations involves terms like $\Delta x^{3}/|\Delta x|$.

For radial separation, we obtain
\begin{align}
\label{eq:psi0radial}
\Psi_{0}^{(1)}(\Delta r)=&\frac{a \Delta r}{\bar{\Delta}^{1/2}(\bar{\Delta}-a^{2}\sin^{2}\bar{\theta})^{1/2}}\nonumber\\
\Psi_{0}^{(2)}(\Delta r)=&\frac{a (\bar{r}-M)(2\bar{\Delta}-a^{2}\sin^{2}\bar{\theta})\Delta r^{2}}{2\bar{\Delta}^{3/2}(\bar{\Delta}-a^{2}\sin^{2}\bar{\theta})^{3/2}}\nonumber\\
\Psi_{0}^{(3)}(\Delta r)=&\frac{a}{24\bar{\Delta}^{5/2}(\bar{\Delta}-a^{2}\sin^{2}\bar{\theta})^{5/2}}\Big[4\bar{\Delta}^{2}(6\bar{\Delta}+8 b^{2})\nonumber\\
&-a^{2}\bar{\Delta}\sin^{2}\bar{\theta}(36 b^{2}+a^{2}+24\bar{\Delta})\nonumber\\
&+3a^{4}\sin^{4}\bar{\theta}(4b^{2}+3\bar{\Delta})\Big]
\end{align}
which along with Eq.(\ref{eq:psizeroexp}) defines $\Psi_{0}$ for radial separation up to $\textrm{O}(\epsilon^{3})$ terms.

Similarly, for $\theta$-separation, we have
\begin{align}
\label{eq:psizerotheta}
\Psi_{0}^{(1)}(\Delta\theta)&=\frac{a\Delta\theta}{(\bar{\Delta}-a^{2}\sin^{2}\bar{\theta})^{1/2}} \nonumber\\
\Psi_{0}^{(2)}(\Delta\theta)&=-\frac{a^{3}\cos\bar{\theta}\sin\bar{\theta}\Delta\theta^{2}}{2(\bar{\Delta}-a^{2}\sin^{2}\bar{\theta})^{3/2}}\nonumber\\
\Psi_{0}^{(3)}(\Delta\theta)&=-\frac{a}{24(\bar{\Delta}-a^{2}\sin^{2}\bar{\theta})^{5/2}}\Big[\bar{\Delta}(\bar{\Delta}+b^{2})\nonumber\\
&+4a^{2}\sin^{2}\bar{\theta}(\bar{\Delta}-2a^{2})+4a^{4}\sin^{4}\bar{\theta}\Big]
\end{align}
while for $\phi$-separation, we have
\begin{align}
\label{eq:psizerophi}
\Psi_{0}^{(1)}(\Delta\phi)&=\frac{\Delta\phi\,\bar{\Delta}^{1/2}}{(\bar{\Delta}^{1/2}-a \,\sin\bar{\theta})}\nonumber\\
\Psi_{0}^{(2)}(\Delta\phi)&=0\nonumber\\
\Psi_{0}^{(3)}(\Delta\phi)&=-\frac{a \,\bar{\Delta}^{1/2}\sin\bar{\theta}(\bar{\Delta}+b^{2}\sin^{2}\bar{\theta})\Delta\phi^{3}}{24(\bar{\Delta}^{1/2}-a\,\sin\bar{\theta})^{4}}.
\end{align}

\subsubsection{The Green's Function for Small Separations}
We now employ these expansions for $\Psi_{0}$ in order to calculate the retarded Green's function expansions for small separations up to $\textrm{O}(\Delta x)$ (or $\textrm{O}(\epsilon)$).

We begin with radial separation. The Green's function expression Eq.(\ref{eq:greensfnstatickerr}) for radial separation is given by
\begin{align}
G_{\textrm{static}}^{r}(r, \bar{r}, \bar{\theta})=\frac{b}{4\pi}\frac{1}{\Delta^{1/2}\bar{\Delta}^{1/2}\sinh[\tfrac{b}{a}\Psi_{0}]-ab\sin^{2}\bar{\theta}\sin\Psi_{0}}
\end{align}
where $\Psi_{0}=\Psi_{0}^{(1)}(\Delta r)\epsilon+\Psi_{0}^{(2)}(\Delta r)\epsilon^{2}+\Psi_{0}^{(3)}(\Delta r)\epsilon^{3}+\textrm{O}(\epsilon^{4})$ with $\Psi_{0}^{(1)}$, $\Psi_{0}^{(2)}$ etc.  given by Eq.(\ref{eq:psi0radial}). Writing $r=\bar{r}-\epsilon \Delta r$ and doing a series expansion about $\epsilon=0$ (this is equivalent to a small $\Delta x$ expansion), we obtain
\begin{widetext}
\begin{align}
\label{eq:greensfnexpradial}
G_{\textrm{static}}^{r}(r, \bar{r}, \bar{\theta})=&\frac{1}{4\pi}\bigg[\frac{\bar{\Delta}^{1/2}}{(\bar{\Delta}-a^{2}\sin^{2}\bar{\theta})^{1/2}\Delta r}+\frac{a^{2}(\bar{r}-M)\sin^{2}\bar{\theta}}{2\bar{\Delta}^{1/2}(\bar{\Delta}-a^{2}\sin^{2}\bar{\theta})^{3/2}}\nonumber\\
& +\frac{a^{2}\sin^{2}\bar{\theta}[\bar{\Delta}(4\bar{\Delta}+4 b^{2}-a^{2})-a^{2}\sin^{2}\bar{\theta}(\bar{\Delta}+2 b^{2})]\Delta r}{8 \bar{\Delta}^{3/2}(\bar{\Delta}-a^{2}\sin^{2}\bar{\theta})^{5/2}}+\textrm{O}(\Delta r^{2})\bigg].
\end{align}
where we have set $\epsilon=1$.

For $\theta$-separation, Eq.(\ref{eq:greensfnstatickerr}) becomes
\begin{align}
G_{\textrm{static}}^{\theta}(\bar{r}, \theta, \bar{\theta})=\frac{b}{4\pi}\frac{1}{\bar{\Delta}\sinh(\tfrac{b}{a}\Psi_{0})-a\, b\sin(\theta-\epsilon\Delta\theta)\,\sin\bar{\theta}\sin\Psi_{0}}
\end{align}
where $\Psi_{0}$ is given by Eq.(\ref{eq:psizeroexp})and (\ref{eq:psizerotheta}). Expanding about $\epsilon=0$ gives
\begin{align}
\label{eq:greensfnexptheta}
G_{\textrm{static}}^{\theta}(\bar{r}, \theta, \bar{\theta})=&\frac{1}{4\pi}\bigg[\frac{1}{(\bar{\Delta}-a^{2}\sin^{2}\bar{\theta})^{1/2}\Delta\theta}-\frac{a^{2}\cos\bar{\theta}\sin\bar{\theta}}{2(\bar{\Delta}-a^{2}\sin^{2}\bar{\theta})^{3/2}}\nonumber\\
&+\frac{[\bar{\Delta}(\bar{\Delta}-3 b^{2})+2 a^{2}\sin^{2}\bar{\theta}(3a^{2}-4\bar{\Delta})-2 a^{4}\sin^{2}\bar{\theta}]\Delta\theta}{24(\bar{\Delta}-a^{2}\sin^{2}\bar{\theta})^{5/2}}+\textrm{O}(\Delta\theta^{2})\bigg].
\end{align}

For $\phi$-separation, the Green's function is
\begin{align}
G_{\textrm{static}}^{\phi}(\bar{r}, \bar{\theta}, \Delta\phi)=\frac{b}{4\pi}\frac{1}{\bar{\Delta}\sinh \big[\tfrac{b}{a}(\Psi_{0}-\Delta\phi)\big]-a\,b \sin^{2}\bar{\theta}\sin\Psi_{0}}
\end{align}
where $\Psi_{0}$ is given by Eq.(\ref{eq:psizeroexp}) and (\ref{eq:psizerophi}). In this case, the series expansion for the Green's function is
\begin{align}
\label{eq:greensfnexpphi}
G_{\textrm{static}}^{\phi}(\bar{r}, \bar{\theta}, \Delta\phi)=&\frac{1}{4\pi}\bigg[\frac{1}{\bar{\Delta}^{1/2}\sin\bar{\theta}\Delta\phi}+\frac{\Delta\phi}{24 \bar{\Delta}^{1/2}(\bar{\Delta}-a^{2}\,\sin^{2}\bar{\theta})^{3}}\Big\{\bar{\Delta}^{3}\csc\bar{\theta}\nonumber\\
&-3 \bar{\Delta}^{2} \sin\bar{\theta}(M^{2}+a^{2})-8 a \,M^{2}\bar{\Delta}^{3/2}\sin^{2}\bar{\theta}\nonumber\\
&-3a^{2}\bar{\Delta}\sin^{3}\bar{\theta}(2M^{2}-a^{2})+a^{4}b^{2}\sin^{5}\bar{\theta}\Big\}+\textrm{O}(\Delta\phi^{2})\bigg].
\end{align}

\subsubsection{Analytic Expression for the Self-Force}
We now have expansions for both the Green's function and the singular field required to compute the self-force. As mentioned above, a particular component of the self-force may be evaluated by taking partial coincidence limits inside the derivative. For example, the radial component is given by
\begin{equation}
f_{r}^{\textrm{self}}=4\pi q^{2}\lim_{r\rightarrow \bar{r}}\bigg[\sqrt{-\bar{g}_{tt}}\,\,\partial_{r}\left(G_{\textrm{static}}^{r}(r, \bar{r},\bar{\theta})-G_{\textrm{sing}}^{r}(r, \bar{r},\bar{\theta})\right)\bigg],
\end{equation}
where $G_{\textrm{static}}^{r}(r, \bar{r},\bar{\theta})$ and $G_{\textrm{sing}}^{r}(r, \bar{r},\bar{\theta})$ are the radially separated Green's function and singular field, respectively. We have computed the Green's function for the various separations in the previous section, the singular field for a particular separation is given by Eq.(\ref{eq:GsingPQR}) in the partial coincidence limit. After some algebra, we obtain
\begin{align}
\label{eq:gsingradial}
G_{\textrm{sing}}^{r}(r, \bar{r},\bar{\theta})=&\frac{1}{4\pi}\bigg[\frac{\bar{\Delta}^{1/2}}{(\bar{\Delta}-a^{2}\sin^{2}\bar{\theta})^{1/2}\Delta r}+\frac{a^{2}(\bar{r}-M)\sin^{2}\bar{\theta}}{2\bar{\Delta}^{1/2}(\bar{\Delta}-a^{2}\sin^{2}\bar{\theta})^{3/2}} \nonumber\\
&+\frac{a^{2}\sin^{2}\bar{\theta}[\bar{\Delta}(4\bar{\Delta}+4 b^{2}-a^{2})-a^{2}\sin^{2}\bar{\theta}(\bar{\Delta}+2 b^{2})]\Delta r}{8 \bar{\Delta}^{3/2}(\bar{\Delta}-a^{2}\sin^{2}\bar{\theta})^{5/2}}+\textrm{O}(\Delta r^{2})\bigg],\\
\label{eq:gsingtheta}
G_{\textrm{sing}}^{\theta}(\bar{r},\theta, \bar{\theta})=&\frac{1}{4\pi}\bigg[\frac{1}{(\bar{\Delta}-a^{2}\sin^{2}\bar{\theta})^{1/2}\Delta\theta}-\frac{a^{2}\cos\bar{\theta}\sin\bar{\theta}}{2(\bar{\Delta}-a^{2}\sin^{2}\bar{\theta})^{3/2}}\nonumber\\
&+\frac{[\bar{\Delta}(\bar{\Delta}-3 b^{2})+2 a^{2}\sin^{2}\bar{\theta}(3a^{2}-4\bar{\Delta})-2 a^{4}\sin^{2}\bar{\theta}]\Delta\theta}{24(\bar{\Delta}-a^{2}\sin^{2}\bar{\theta})^{5/2}}+\textrm{O}(\Delta\theta^{2})\bigg],\\
\label{eq:gsingphi}
G_{\textrm{sing}}^{\phi}(\bar{r},\bar{\theta}, \Delta\phi)=&\frac{1}{4\pi}\bigg[\frac{1}{\bar{\Delta}^{1/2}\sin\bar{\theta}\Delta\phi}+\frac{\Delta\phi}{24\bar{\Delta}^{1/2}(\bar{\Delta}-a^{2}\sin^{2}\bar{\theta})^{3}}\Big\{\bar{\Delta}^{3}\csc\bar{\theta}\nonumber\\
&-3 \bar{\Delta}^{2} \sin\bar{\theta}(M^{2}+a^{2})-3a^{2}\bar{\Delta}\sin^{3}\bar{\theta}(2M^{2}-a^{2})+a^{4}b^{2}\sin^{5}\bar{\theta}\Big\}+\textrm{O}(\Delta\phi^{2})\bigg].
\end{align}
\end{widetext}

Comparison of Eq.(\ref{eq:greensfnexpradial}) with Eq.(\ref{eq:gsingradial}), we see that
\begin{equation}
G_{\textrm{static}}^{r}(r,\bar{r},\bar{\theta})-G_{\textrm{sing}}^{r}(r,\bar{r},\bar{\theta})=\textrm{O}(\Delta r^{2})
\end{equation}
and therefore Eq.(\ref{eq:selfforcestaticvacuumexp}) clearly gives
\begin{equation}
\label{eq:selfforcekerrradial}
f^{\textrm{self}}_{r}=0.
\end{equation}
Similarly, from Eqs.(\ref{eq:greensfnexptheta}) and (\ref{eq:gsingtheta}), we have
\begin{equation}
G_{\textrm{static}}^{\theta}(\bar{r},\theta, \bar{\theta})-G_{\textrm{sing}}^{\theta}(\bar{r},\theta, \bar{\theta})=\textrm{O}(\Delta \theta^{2})
\end{equation}
and hence Eq.(\ref{eq:selfforcestaticvacuumexp}) yields
\begin{equation}
\label{eq:selfforcekerrtheta}
f^{\textrm{self}}_{\theta}=0.
\end{equation}
Therefore the only component of the self-force lies in the azimuthal direction, as suggested by the numerical results of Burko and Liu \cite{BurkoLiu}. Subtracting Eq.(\ref{eq:gsingphi}) from Eq.(\ref{eq:greensfnexpphi}) gives the regular field
\begin{align}
&G_{\textrm{static}}^{\phi}(\bar{r},\bar{\theta}, \Delta\phi)-G_{\textrm{sing}}^{\phi}(\bar{r},\bar{\theta}, \Delta\phi)\nonumber\\
&=-\frac{a\,M^{2}\bar{\Delta}\,\sin^{2}\bar{\theta}\Delta\phi}{12\pi\, (\bar{\Delta}-a^{2}\sin^{2}\bar{\theta})^{3}}+\textrm{O}(\Delta\phi^{2}).
\end{align}
Differentiating with respect to $\phi$, including the $4\pi/\sqrt{-\bar{g}_{tt}}$ term and taking the limit $\phi\rightarrow\bar{\phi}$ according to Eq.(\ref{eq:selfforcestaticvacuumexp}) gives the azimuthal component of the self-force on a static particle in the Kerr space-time:
\begin{align}
\label{eq:selfforcekerrphi}
f_{\phi}^{\textrm{self}}(\bar{x})=\frac{1}{3}q^{2}a\,\bar{\Delta} \,\sin^{2}\bar{\theta}\frac{M^{2}}{(\bar{\Delta}-a^{2}\,\sin^{2}\bar{\theta})^{5/2}\bar{\Sigma}^{1/2}}.
\end{align}
Our results clearly agree with the numerical results of \cite{BurkoLiu}. We infer from this agreement that the MSRP parameter $D_{a}$ that the authors of Ref.~\cite{BurkoLiu} conjecture is correct, at least for a static particle. Our analytic results ought to offer a good standard test for any numerical regularization prescription in the non-spherically symmetric case. We note also that the result presented here is trivially extended to the Kerr-Newman black hole by taking $M^{2}\rightarrow M^{2}-Q^{2}$, where $Q$ is the charge of the black hole.

Combining the results Eqs.(\ref{eq:selfforcekerrradial}), (\ref{eq:selfforcekerrtheta}) and (\ref{eq:selfforcekerrphi}) and assuming, for notational convenience, that the scalar charge is located at Boyer-Lindquist coordinates $(r, \theta, \phi)$ gives the following representation of the self-force on a static scalar charge in the Kerr background geometry:
\begin{align}
f_{b}^{\textrm{self}}=\frac{1}{3}q^{2}a\,\Delta \,\sin^{2}\theta\frac{M^{2}}{(\Delta-a^{2}\,\sin^{2}\theta)^{5/2}\Sigma^{1/2}}\delta^{\phi}{}_{b}.
\end{align}

\section{Conclusions}
In this paper, we derived a completely closed-form representation of the retarded Green's function for a scalar particle at fixed spatial Boyer-Lindquist coordinates in the Kerr space-time. This representation relied on two results for the Associated Legendre functions, one of which is derived in the Appendix and the other we derive elsewhere \cite{CasalsOttewillTaylor}. These formulae will likely prove useful in other contexts. As a check of the validity of our closed-form Green's function, we computed, analytically, the self-force on a static scalar particle in Kerr space-time and showed that our result agrees with the conjectured result of Burko and Liu \cite{BurkoLiu}. The calculation presented in this paper is therefore the first proof of the result implied by the numerical work of Ref.~\cite{BurkoLiu}.


\appendix
\section*{Appendix}
\subsection*{Summation Formula for the Product of Associated Legendre Functions}
\label{app:LegendreProduct}
In this appendix, we shall derive the following summation formula for Associated Legendre functions:
\begin{align}
\label{eq:baranovsum}
\sum_{l=|m|}^{\infty}(2l+1)\frac{(l-m)!}{(l+m)!}P_{l}^{m}(\cos\theta)P_{l}^{m}(\cos\theta')P_{l}(x)\nonumber\\
=\frac{2}{\pi}\frac{\cos(m\cos^{-1}(\zeta))}{(\sin^{2}\theta\sin^{2}\theta'-(x-\cos\theta\cos\theta')^{2})^{1/2}}
\end{align}
where
\begin{equation}
\zeta=\frac{x-\cos\theta\cos\theta'}{\sin\theta\sin\theta'}
\end{equation}
and $x=\cos\lambda$ must lie in the range $\theta-\theta'<\lambda<\min\{\theta+\theta', 2\pi-\theta-\theta'\}$. For $\lambda$ outside of this range, the series sums to zero. Though not a new result, it is not well known, the only reference of which we are aware that includes this formula is Ref.~\cite{Hansen}. We have therefore included a proof of this result, following closely the method of Baranov \cite{Baranov} who has proven the $m=0$ case, as the technique might be valuably extended in other directions.
We begin with the well-known Christoffel-Darboux formula \cite{Erdelyi}
\begin{align}
&\sum_{l=0}^{n} \frac{(2 l+1)}{2}P_{l}(t)P_{l}(x)\nonumber\\
&=\frac{n+1}{2}\frac{P_{n+1}(x)P_{n}(t)-P_{n}(x)P_{n+1}(t)}{x-t}.
\end{align}
Taking
\begin{align}
x&=\cos\lambda,\nonumber\\
 t&=\cos\mu=\cos\theta\cos\theta'+\sin\theta\sin\theta'\cos\varphi,
\end{align}
we may employ the Legendre addition theorem to obtain
\begin{widetext}
\begin{align}
\sum_{l=0}^{n}\sum_{m=-l}^{l}\cos(m\varphi) \frac{(2 l+1)}{2}\frac{(l-m)!}{(l+m)!}P_{l}^{m}(\cos\theta)P_{l}^{m}(\cos\theta')P_{l}(x)=\frac{n+1}{2}\frac{P_{n+1}(x)P_{n}(t)-P_{n}(x)P_{n+1}(t)}{x-t}.
\end{align}
Reversing the order of summation, multiplying across by $\cos(m' \varphi)$ and integrating from $0$ to $2\pi$ gives
\begin{align}
\label{eq:sumtonintphi}
\pi \sum_{l=|m|}^{n}(2l+1)\frac{(l-m)!}{(l+m)!}P_{l}^{m}(\cos\theta)P_{l}^{m}(\cos\theta')P_{l}(x)=\frac{n+1}{2}\int_{0}^{2\pi}\cos(m\varphi)\frac{P_{n+1}(x)P_{n}(t)-P_{n}(x)P_{n+1}(t)}{x-t}d\varphi.
\end{align}
We change the integration parameter from $\varphi$ to $t$ using
\begin{align}
d\varphi=\pm\frac{dt}{\sqrt{(\cos(\theta-\theta')-t)(t-\cos(\theta+\theta'))}}
\end{align}
where the choice of sign here is determined by the fact that $t$, as a function of $\varphi$, is symmetrical about $\varphi=\pi$, decreasing from $\cos(\theta-\theta')$ at $\varphi=0$ to $\cos(\theta+\theta')$ at $\varphi=\pi$, and then increasing symmetrically to $\cos(\theta-\theta')$ at $\varphi=2\pi$. Hence, we obtain
\begin{align}
\label{eq:sumtonintt}
&\pi\sum_{l=|m|}^{n}(2l+1)\frac{(l-m)!}{(l+m)!}P_{l}^{m}(\cos\theta)P_{l}^{m}(\cos\theta')P_{l}(x)\nonumber\\
&=(n+1)\int_{\cos(\theta+\theta')}^{\cos(\theta-\theta')}\cos(m\,\varphi(t))\frac{P_{n+1}(x)P_{n}(t)-P_{n}(x)P_{n+1}(t)}{(x-t)\sqrt{(\cos(\theta-\theta')-t)(t-\cos(\theta+\theta'))}}dt
\end{align}
where
\begin{equation}
\varphi(t)=\cos^{-1}\Big(\frac{t-\cos\theta\cos\theta'}{\sin\theta\sin\theta'}\Big).
\end{equation}
In order to obtain the result (\ref{eq:baranovsum}), we require an asymptotic analysis of the integral in Eq.(\ref{eq:sumtonintt}) for large $n$ so that the limit as $n\rightarrow \infty$ may be taken. Such an asymptotic analysis is complicated further by the fact that one needs to allow for the possibility of small or vanishing $x-t$ (which only occurs when $x$ lies inside the integration range). Using the large $n$ asymptotic formula \cite{gradriz}
\begin{equation}
\label{eq:plargen}
P_{n}(\cos z)=\Big(\frac{2}{\pi n \sin z}\Big)^{1/2}\cos[(n+\tfrac{1}{2})z-\tfrac{\pi}{4}]-\frac{\pi}{16}\Big(\frac{2}{\pi n \sin z}\Big)^{3/2}\cos[(n+\tfrac{3}{2})z+\tfrac{\pi}{4}]+\textrm{O}(n^{-5/2}),
\end{equation}
we can deduce
\begin{align}
\label{eq:christoffellargen}
P_{n+1}(x)P_{n}(t)-P_{n}(x)P_{n+1}(t)=\frac{2[\sin(\tfrac{1}{2}(\lambda-\mu))\cos((n+1)(\lambda+\mu))-\sin(\tfrac{1}{2}(\lambda+\mu))\sin((n+1)(\lambda-\mu))]}{\pi(n+1)\sqrt{\sin\mu \sin\lambda}}+A_{n}(x,t)
\end{align}
where $t=\cos\mu$, $x=\cos\lambda$ and the remainder $A_{n}(x,t)$ is given by
\begin{align}
A_{n}(x,t)=&\frac{1}{4\pi n^{2}}\bigg(\frac{\cos[(n+\tfrac{1}{2})\lambda-\tfrac{\pi}{4}]\cos[(n+\tfrac{5}{2})\mu+\tfrac{\pi}{4}]}{\sqrt{\sin\lambda \sin^{3}\mu}}-\frac{\cos[(n+\tfrac{3}{2})\lambda-\tfrac{\pi}{4}]\cos[(n+\tfrac{3}{2})\mu+\tfrac{\pi}{4}]}{\sqrt{\sin\lambda \sin^{3}\mu}}\nonumber\\
&+\frac{\cos[(n+\tfrac{3}{2})\lambda+\tfrac{\pi}{4}]\cos[(n+\tfrac{3}{2})\mu-\tfrac{\pi}{4}]}{\sqrt{\sin^{3}\lambda \sin\mu}}-\frac{\cos[(n+\tfrac{5}{2})\lambda+\tfrac{\pi}{4}]\cos[(n+\tfrac{1}{2})\mu-\tfrac{\pi}{4}]}{\sqrt{\sin^{3}\lambda \sin\mu}}\bigg)+\textrm{O}(n^{-3}).
\end{align}
One can choose a function of $x$ such that the trigonometric terms in the brackets are bounded for all values of $t$, i.e.
\begin{equation}
\label{eq:remainder1est}
|A_{n}(x,t)|<\frac{c(x)}{n^{2}}
\end{equation}
where $c(x)$ is independent of $n$ and $t$. The approximation (\ref{eq:christoffellargen}) with Eq.(\ref{eq:remainder1est}) is valid whenever $x$ is not close to $t$ but as mentioned above, we require an alternative estimate to that of Eq.(\ref{eq:remainder1est}) that accounts for small or vanishing $x-t$. To achieve this, we can use the asymptotic formula for the associated Legendre function of order $1$
\begin{equation}
\label{eq:p1largen}
P_{l}^{1}(\cos z)=\Big(\frac{2 n}{\pi \sin z}\Big)^{1/2}\cos[(n+\tfrac{1}{2})z+\tfrac{\pi}{4}]+\frac{3\pi}{16 n^{2}}\Big(\frac{2 n}{\pi \sin z}\Big)^{3/2}\cos[(n+\tfrac{3}{2})z+\tfrac{3\pi}{4}]+\textrm{O}(n^{-3/2}),
\end{equation}
as well as Eq.(\ref{eq:plargen}) to obtain the estimate
\begin{align}
\label{eq:christoffeldifflargen}
P_{n}(x)P_{n+1}^{1}(t)-P_{n+1}(x)P_{n}^{1}(t)=\frac{2[\sin(\tfrac{1}{2}(\lambda-\mu))\sin((n+1)(\lambda+\mu))-\sin(\tfrac{1}{2}(\lambda+\mu))\cos((n+1)(\lambda-\mu))]}{\pi\sqrt{\sin\mu \sin\lambda}}+B_{n}(x,t),
\end{align}
where
\begin{align}
B_{n}(x,t)=&\frac{1}{4\pi n}\bigg(\frac{3\cos[(n+\tfrac{1}{2})\lambda-\tfrac{\pi}{4}]\cos[(n+\tfrac{5}{2})\mu+\tfrac{3\pi}{4}]}{\sqrt{\sin\lambda \sin^{3}\mu}}-\frac{3\cos[(n+\tfrac{3}{2})\lambda-\tfrac{\pi}{4}]\cos[(n+\tfrac{3}{2})\mu+\tfrac{3\pi}{4}]}{\sqrt{\sin\lambda \sin^{3}\mu}}\nonumber\\
&+\frac{\cos[(n+\tfrac{5}{2})\lambda+\tfrac{\pi}{4}]\cos[(n+\tfrac{1}{2})\mu+\tfrac{\pi}{4}]}{\sqrt{\sin^{3}\lambda \sin\mu}}-\frac{\cos[(n+\tfrac{3}{2})\lambda+\tfrac{\pi}{4}]\cos[(n+\tfrac{3}{2})\mu+\tfrac{\pi}{4}]}{\sqrt{\sin^{3}\lambda \sin\mu}}\bigg)+\textrm{O}(n^{-2}).
\end{align}
Given that
\begin{equation}
P_{l}^{1}(t)=-\sin\mu \frac{d P_{l}(t)}{d t},
\end{equation}
we may divide both sides of Eq.(\ref{eq:christoffeldifflargen}) by $\sin\mu$ and integrate from $x$ to $t$ to get
\begin{align}
P_{n+1}(x)P_{n}(t)-P_{n}(x)P_{n+1}(t)=&\frac{2[\sin(\tfrac{1}{2}(\lambda-\mu))\cos((n+1)(\lambda+\mu))-\sin(\tfrac{1}{2}(\lambda+\mu))\sin((n+1)(\lambda-\mu))]}{\pi(n+1)\sqrt{\sin\mu \sin\lambda}}\nonumber\\
&+\int_{x}^{t}\frac{B_{n}(x,t)}{\sin\mu}dt.
\end{align}
Comparing with (\ref{eq:christoffellargen}), we see that
\begin{equation}
A_{n}(x,t)=\int_{x}^{t}\frac{B_{n}(x,t)}{\sin\mu}dt.
\end{equation}
Again the trigonometric terms in $B_{n}(x,t)/\sin\mu$ are bounded. If we take $m$ to be the lower bound and $M$ the upper bound and define the positive constant $c_{1}=|\max\{m,M\}|$, then we get the following bound for $A_{n}$
\begin{equation}
\label{eq:remainder2est}
|A_{n}(x,t)|=\Big|\int_{x}^{t}\frac{B_{n}(x,t)}{\sin\mu}dt\Big|<\frac{c_{1}|x-t|}{n}.
\end{equation}
The geometric mean of Eq.(\ref{eq:remainder1est}) and (\ref{eq:remainder2est}) yields a new estimate for the remainder term:
\begin{equation}
\label{eq:remaindermean}
|A_{n}(x,t)|<\frac{\sqrt{c(x) c_{1}|x-t|}}{n^{3/2}}.
\end{equation}
Employing Eq.(\ref{eq:christoffellargen}) with the estimate (\ref{eq:remaindermean}), we have
\begin{align}
&\pi\sum_{l=|m|}^{n}(2l+1)\frac{(l-m)!}{(l+m)!}P_{l}^{m}(\cos\theta)P_{l}^{m}(\cos\theta')P_{l}(x)=\nonumber\\
&\int_{\cos(\theta+\theta')}^{\cos(\theta-\theta')}\frac{2\cos(m\,\varphi(t))}{\pi(x-t)\sqrt{\sin\mu \sin\lambda}\sqrt{(\cos(\theta-\theta')-t)(t-\cos(\theta+\theta'))}}\nonumber\\
&\times\Big[\sin(\tfrac{1}{2}(\lambda-\mu))\cos((n+1)(\lambda+\mu))-\sin(\tfrac{1}{2}(\lambda+\mu))\sin((n+1)(\lambda-\mu))\Big]dt+\mathrm{O}(\frac{1}{\sqrt{n}}).
\end{align}
It is convenient to integrate over $\mu$ rather than $t$, and express the singularity as
\begin{equation}
\frac{1}{x-t}=\frac{1}{\cos\lambda-\cos\mu}=\frac{1}{(\mu-\lambda)\sin\lambda}+f(\lambda,\mu)
\end{equation}
where $f$ is a continuous function. Since $f$ contains no singularities, and in view of Eq.(\ref{eq:remainder1est}), it is easy to see that the contribution of the integral containing $f$ in the integrand is $\mathrm{O}(n^{-1})$. Therefore,
\begin{align}
\label{eq:sumtonintmu}
&\pi\sum_{l=|m|}^{n}(2l+1)\frac{(l-m)!}{(l+m)!}P_{l}^{m}(\cos\theta)P_{l}^{m}(\cos\theta')P_{l}(x)=\nonumber\\
&\int_{\mu_{-}}^{\mu_{+}}\frac{2\cos(m\,\varphi(\mu))}{\pi\sqrt{(\cos(\theta-\theta')-\cos\mu)(\cos\mu-\cos(\theta+\theta'))}}\sqrt{\frac{\sin\mu}{\sin^{3}\lambda}}\nonumber\\
&\times\Big[\frac{\sin(\tfrac{1}{2}(\lambda-\mu))\cos((n+1)(\lambda+\mu))-\sin(\tfrac{1}{2}(\lambda+\mu))\sin((n+1)(\lambda-\mu))}{\mu-\lambda}\Big]d\mu+\mathrm{O}(\frac{1}{\sqrt{n}}).
\end{align}
where $\mu_{-}=\theta-\theta'$ and $\mu_{+}=\min\{\theta+\theta',2\pi-\theta-\theta'\}$. Without loss of generality, in what follows, we shall assume that $\theta+\theta'<\pi$ so that $\mu_{+}=\theta+\theta'$.

The first integral in Eq.(\ref{eq:sumtonintmu}) may be written as
\begin{align}
\int_{\mu_{-}}^{\mu_{+}}\cos(m\,\varphi(\mu))\frac{2\,\sin(\tfrac{1}{2}(\lambda-\mu))\cos((n+1)(\lambda+\mu))}{\pi\sqrt{(\cos(\theta-\theta')-\cos\mu)(\cos\mu-\cos(\theta+\theta'))}}\sqrt{\frac{\sin\mu}{\sin^{3}\lambda}}\frac{d\mu}{\mu-\lambda}=\int_{\mu_{-}}^{\mu_{+}}q_{m}(\mu)\cos((n+1)(\lambda+\mu))d\mu
\end{align}
where $q_{m}$ is independent of $n$ but not of $m$. One can further write
\begin{equation}
\label{eq:q}
q_{m}(\mu)=\frac{A_{m}}{\sqrt{\mu-(\theta-\theta')}}+\frac{B_{m}}{\sqrt{\theta+\theta'-\mu}}+q_{m,1}(\mu)
\end{equation}
for particular constants $A_{m}$ and $B_{m}$ and some continuous function $q_{m,1}(\mu)$. Considering the first singular term here:
\begin{align}
\int_{\mu_{-}}^{\mu_{+}}\frac{\cos((n+1)(\lambda+\mu))}{\sqrt{\mu-(\theta-\theta')}}d\mu=& 2\int_{0}^{\sqrt{2 \theta'}}\cos((n+1)(\lambda+\theta-\theta'+s^{2}))ds \nonumber\\
=& 2\cos[(n+1)(\lambda+\theta-\theta')]\int_{0}^{\sqrt{2\theta'}}\cos[(n+1)s^{2}]ds\nonumber\\
& -2\sin[(n+1)(\lambda+\theta-\theta')]\int_{0}^{\sqrt{2\theta'}}\sin[(n+1)s^{2}]ds
\end{align}
where we adopted the transformation $s=\sqrt{\mu-(\theta-\theta')}$ in the first line. We further make the transformation $S=s\sqrt{n+1} $ which gives
\begin{align}
\int_{\mu_{-}}^{\mu_{+}}\frac{\cos((n+1)(\lambda+\mu))}{\sqrt{\mu-(\theta-\theta')}}d\mu=&\frac{2\cos[(n+1)(\lambda+\theta-\theta')]}{\sqrt{n+1}}\int_{0}^{\sqrt{2(n+1)\theta'}}\cos (S^{2}) dS\nonumber\\
&-\frac{2\sin[(n+1)(\lambda+\theta-\theta')]}{\sqrt{n+1}}\int_{0}^{\sqrt{2(n+1)\theta'}}\sin (S^{2}) dS.
\end{align}
These last two integral are the well-known Fresnel integrals and they converge for large values of the argument ($n\rightarrow \infty$). Since $\cos(n x)/\sqrt{n} \rightarrow 0$ and $\sin(n x)/\sqrt{n}\rightarrow 0$ as $n\rightarrow \infty$, we must have
\begin{align}
\int_{\mu_{-}}^{\mu_{+}}\frac{\cos((n+1)(\lambda+\mu))}{\sqrt{\mu-(\theta-\theta')}}d\mu\rightarrow 0\qquad \textrm{as}\qquad n\rightarrow\infty.
\end{align}
An identical analysis for the other singular term in Eq.(\ref{eq:q}) reveals
\begin{align}
\int_{\mu_{-}}^{\mu_{+}}\frac{\cos((n+1)(\lambda+\mu))}{\sqrt{\theta+\theta'-\mu}}d\mu=
\rightarrow\,\, 0\qquad \textrm{as}\qquad n\rightarrow \infty.
\end{align}
Hence, the contributions from the singular terms in Eq.(\ref{eq:q}) vanish for large $n$.

Turning our attention now to the regular part of Eq.(\ref{eq:q}), taking our integral over the entire real line (since consideration of Eqs.(\ref{eq:sumtonintt}), (\ref{eq:christoffellargen}) and (\ref{eq:remainder1est}) show that $q_{m,1}(\mu)$ is $\textrm{O}(n^{-1})$ outside the range of integration and therefore does not contribute in the large $n$ limit), we obtain
\begin{align}
\int_{-\infty}^{\infty}q_{m,1}(\mu)\cos((n+1)(\lambda+\mu))d\mu=\frac{1}{2}\int_{-\infty}^{\infty}q_{m,1}(\mu)\cos((n+1)(\lambda+\mu))d\mu+\frac{1}{2}\int_{-\infty}^{\infty}q_{m,1}(\nu)\cos((n+1)(\lambda+\nu))d\nu.
\end{align}
Making the substitution
\begin{equation}
\nu=\mu-\frac{\pi}{n+1}
\end{equation}
in the second integral yields
\begin{align}
\int_{-\infty}^{\infty}q_{m,1}(\mu)\cos((n+1)(\lambda+\mu))d\mu=\frac{1}{2}\int_{-\infty}^{\infty}\Big[q_{m,1}(\mu)-q_{m,1}\Big(\mu-\frac{\pi}{n+1}\Big)\Big]\cos((n+1)(\lambda+\mu))d\mu
\end{align}
which vanishes as $n\rightarrow \infty$.

Finally, we consider the second integral in Eq.(\ref{eq:sumtonintmu}). This integral may be written as
\begin{align}
\label{eq:intt}
\int_{\mu_{-}}^{\mu_{+}}\frac{\sin((n+1)(\mu-\lambda))}{\mu-\lambda}T_{m}(\mu) d\mu=\int_{\mu_{-}}^{\mu_{+}}\{\sin((n+1)(\mu-\lambda))T_{m}(\lambda)+\sin((n+1)(\mu-\lambda))[T_{m}(\mu)-T_{m}(\lambda)]\}\frac{d\mu}{\mu-\lambda}
\end{align}
where
\begin{equation}
T_{m}(\mu)=\frac{2}{\pi}\frac{\cos(m\,\varphi(\mu))\sin(\tfrac{1}{2}(\lambda+\mu)}{\sqrt{(\cos(\theta-\theta')-\cos\mu)(\cos\mu-\cos(\theta+\theta'))}}\sqrt{\frac{\sin\mu}{\sin^{3}\lambda}}.
\end{equation}
For the second integral in the expression above, one can write
\begin{equation}
\frac{T_{m}(\mu)-T_{m}(\lambda)}{\mu-\lambda}=\frac{C_{m}}{\sqrt{\mu-(\theta-\theta')}}+\frac{D_{m}}{\sqrt{\theta+\theta'-\mu}}+p_{m}(\mu)
\end{equation}
where $p_{m}$ is a continuous function. Applying a similar analysis as that above, one can show that the contribution coming from this integral converges to zero as $n\rightarrow \infty$.

The only contribution to the sum in (\ref{eq:sumtonintmu}) therefore comes from the first integral on the right-hand side of (\ref{eq:intt}):
\begin{equation}
T_{m}(\lambda)\int_{\theta-\theta'}^{\theta+\theta'}\frac{\sin((n+1)(\mu-\lambda))}{\mu-\lambda} d\mu=T_{m}(\lambda)\int_{(n+1)(\theta-\theta'-\lambda)}^{(n+1)(\theta+\theta'-\lambda)} \frac{\sin\tau}{\tau} d\tau .
\end{equation}
In the limit as $n\rightarrow \infty$, this integral converges to the Dirichlet integral,
\begin{align}
\lim_{n\rightarrow\infty}\Big[T_{m}(\lambda)\int_{(n+1)(\theta-\theta'-\lambda)}^{(n+1)(\theta+\theta'-\lambda)} \frac{\sin\tau}{\tau} d\tau\Big]&=T_{m}(\lambda)\int_{-\infty}^{\infty}\frac{\sin\tau}{\tau}d\tau = \pi T_{m}(\lambda) \nonumber\\
&=\frac{2 \cos(m\, \varphi(\lambda))}{\sqrt{\sin^{2}\theta\sin^{2}\theta'-(\cos\lambda-\cos\theta\cos\theta')^{2}}}
\end{align}
where
\begin{equation}
\varphi(\lambda)=\cos^{-1}\Big(\frac{\cos\lambda-\cos\theta\cos\theta'}{\sin\theta\sin\theta'}\Big).
\end{equation}
We note here that for $\lambda$ outside the range $(\mu_{-},\mu_{+})$, the integral (\ref{eq:intt}) would have no singularities and the integral would have vanished in the limit $n\rightarrow \infty$. Therefore we have shown
\begin{equation} 
\sum_{l=|m|}^{\infty}(2l+1)\frac{(l-m)!}{(l+m)!}P_{l}^{m}(\cos\theta)P_{l}^{m}(\cos\theta')P_{l}(x)=\frac{2}{\pi}\frac{\cos(m\cos^{-1}(\zeta))}{(\sin^{2}\theta\sin^{2}\theta'-(x-\cos\theta\cos\theta')^{2})^{1/2}}
\end{equation}
where
\begin{equation}
\zeta=\frac{x-\cos\theta\cos\theta'}{\sin\theta\sin\theta'}
\end{equation}
and $x=\cos\lambda$ must lie in the range $\theta-\theta'<\lambda<\min\{\theta+\theta', 2\pi-\theta-\theta'\}$.
\end{widetext}

\bibliographystyle{apsrev}
\bibliography{database}

\end{document}